\documentclass[twocolumn]{aastex631}

\shorttitle{Computational Models of Three Kepler Binaries}
\shortauthors{Odesse et al.}
\graphicspath{{./}{figures/}}

\begin{document}

\title{Using Computational Models to Uncover the Parameters of Three \textit{Kepler} Binaries: KIC~5957123, KIC~8314879, and KIC~10727668\footnote{Released on August, 20th, 2021}}

\author{Padraic E. Odesse}
\affiliation{McMaster University \\
1280 Main Street West \\
Hamilton, Ontario}
\affiliation{Mount Allsion University \\
62 York Street \\
Sackville, New Brunswick}

\author[0000-0003-4233-3105]{Catherine Lovekin}
\affiliation{Mount Allsion University \\
62 York Street \\
Sackville, New Brunswick}



\begin{abstract}
Theories of stellar convective core overshoot can be examined through analysis of pulsating stars. Better accuracy can be achieved by obtaining external constraints such as those provided by observing pulsating stars in eclipsing binary systems, but this requires that the binary parameters be identified so photometric variations of the pulsating component may be isolated from the binary periodicity. 
This study aims to uncover the physical parameters of three binaries observed by the \textit{Kepler} spacecraft. We also seek to evaluate the feasibility of accurately constraining binaries using only readily available time-series photometry and distance estimates.
Binary models were constructed using the Physics of Eclipsing Binaries (PHOEBE) software package. Markov Chain Monte Carlo methods were used to sample the parameter space of these models and provide estimates of the posterior distributions for these systems. An initial run using binned light curve data was performed to identify general parameter trends and provide initializing distributions for a subsequent analysis incorporating the full data set.
We present theoretical models for all three binaries, along with posterior distributions from our MCMC analyses. Models for KIC 8314879 and KIC 10727668 produced a good match to the observed data, while the model of KIC 5957123 failed to generate an appropriate synthetic light curve.
For the two successful models, we interpret the posterior distributions and discuss confidence in our parameter estimates and uncertainties. We also evaluate the feasibility of this procedure in various contexts, and propose several modifications to improve the success of future studies.
\end{abstract}

\keywords{Ellipsoidal Variable Stars (455) --- Close Binary Stars (254) --- Eclipsing Binary Stars (444) --- White Dwarf Stars (1799) --- Markov chain Monte Carlo (1889) --- Astronomical Simulations (1857)}


\section{Introduction} \label{sec:intro}


Eclipsing binary star systems are ideal candidates for testing theoretical stellar models. Using photometric observations, the parameters of eclipsing binaries can be measured to a high degree of precision, which enables us to test theories and make predictions with tight constraints and greater certainty \citep{southworth_binary_2020}. Past research into stellar core convection suggests that there must be some overshoot beyond the theoretical convective boundary \citep{lovekin_convection_2017, pedersen_shape_2018, johnston_one_2021}, but the behaviour of overshoot as a function of stellar characteristics such as mass, age, or metallicity, has not yet been fully characterized. Since overshoot can be studied by examining pulsation modes through astroseismology, any star for which we can perform pulsation analysis provides insight into how overshoot varies. Therefore, identifying a binary system in which pulsation modes are present would provide a valuable target for accurately probing stellar core convection.



Recent exoplanet surveys such as \textit{Kepler} \citep{Borcuki2004}and TESS \citep{Ricker2015} measure variations in stellar flux with very high precision, and as a result are also used to study variable stars.
\textit{Kepler} data is open-access, and over 2800 binary stars have been identified in the \textit{Kepler} catalogue \citep{kirk_keplereclipsing_2016}. If we could identify binary stars with pulsating components from the \textit{Kepler} catalogue, and model them to obtain parameter constraints and extract light curve residuals, then we could better probe the behaviour of convective core overshoot. 

Generally, radial velocity measurements are necessary to fully constrain the parameters of eclipsing binary systems \citep{southworth_binary_2020}. However, radial velocities are derived from stellar spectra, which \textit{Kepler} does not observe. Obtaining spectra for \textit{Kepler} binaries through ground based observations is difficult; many of these binaries are faint (14th to 16th magnitude) and so require long observation times. For instance, \citet{rappaport_discovery_2015} report 15-30 second exposure times at 11 epochs over five months to analyze one \textit{Kepler} binary. 
For this research, we did not have access to an adequate spectrograph or telescope to conduct our own observations of these systems.
A method of constraining stellar parameters using only time-series photometry would be undoubtedly useful, since the observing time required to follow up all of these systems with ground based spectroscopic observations is prohibitive. In this paper, we hope to evaluate the feasibility of constraining binary system and stellar parameters without stellar spectra, instead using detailed stellar models and marginalizing over appropriate parameters with numerical methods.

Theoretical binary models are quite adept at identifying the light curve produced by a binary system with a given set of physical characteristics \citep{prsa_computational_2005}. The inverse of this problem -- identifying the parameter values which best describe a binary system that produced a given light curve -- is more difficult to solve, but we can attempt to do so by applying various numerical methods to our theoretical binary models \citep{conroy_physics_2020}. Generally, this involves creating an ensemble of binary models with varying parameter values, generating a light curve for each model, and comparing it to the observed photometric data using a $\chi^2$ evaluation. We can begin to estimate the physical parameters of this binary by identifying which model parameters yield synthetic observables that best match the photometric data. 
Markov Chain Monte Carlo (MCMC) analysis is used to infer which distribution of parameter values best describe the observed photometric data. 

MCMC methods sample the posterior probability of parameter combinations. That is to say, these methods evaluate the probability that a given parameter combination actually generates the observed data. Performing MCMC ensemble sampling yields an estimate of the posterior density function for the parameter space, which provide regions of confidence around our parameter estimates. Although the posterior distributions do not directly correspond to parameter uncertainties (see \citet{hogg_data_2010} for a mathematical overview, or \citet{maxted_tess_2020} who acknowledge this issue in the context of eclipsing binaries), the results of MCMC sampling provide insight into parameter correlations and, if used correctly, are a key step towards identifying expectation values and uncertainties for our parameter estimates.



In this work, we investigate three binary systems from the Kepler Input Catalogue: KIC 5957123, KIC 8314879, and KIC 10727668. The Fourier spectra of KIC 8314879 and 10727668 show additional frequencies that suggest these are pulsating stars, as well as eclipsing binaries. Section \ref{sec:data} describes the data obtained from the Kepler mission, and the steps taken to prepare this data for analysis. Section \ref{sec:models} describes the computational models of each binary, and Section \ref{sec:mcmc} describes the numerical methods used to evaluate our parameter fits. Section \ref{sec:results} presents the resulting parameter estimates with uncertainties, followed by an evaluation of our models and their limitations in Section \ref{sec:discussion}. We provide a brief summary of our conclusions in Section \ref{sec:conclusions}.

\section{Data} \label{sec:data}

\subsection{The Kepler Input Catalogue} \label{subsec:kepler}

The three systems discussed in this paper were observed by the \textit{Kepler} spacecraft with a cadence of 29.4 min. Four quarters of data are available for each of the three binaries, culminating in 316.8 days worth of observations and roughly 13,500 data points from which a light curve was constructed for each system.

\subsection{Flux Calibration} \label{subsec:calibration}


To allow for proper modeling with PHOEBE, it was necessary to calibrate the differential photometry of the \textit{Kepler} data. Pre-flight estimates suggest that flux from a 12th magnitude solar-type star incident on one of \textit{Kepler}'s CCDs releases a current of $1.74\times10^5$ electrons per second \citep{farmer_true_2013}. 

The \textit{Kepler} magnitude system is based on the AB magnitude system \citep{brown_kepler_2011}. Equation 4 from \citet{tonry_pan-starrs1_2012} may be used to compute the magnitude ($m_{\text{AB}}$) of the Sun through the \textit{Kepler} passband: 
\begin{equation}
	\label{ab_mag}
	m_{\text{AB}}=-2.5\log_{10}\bigg(\frac{\int f_\nu(h\nu)^{-1}A(\nu)d\nu }{\int 3631\text{Jy}(h\nu)^{-1}A(\nu)d\nu }\bigg)
\end{equation}
where $A(\nu)$ is the \textit{Kepler} passband transmission function taken from the \textit{Kepler Instrument Handbook} supplemental \citep{van_cleve_kepler_2009}, and $f_\nu$ is the spectral flux density of the Sun, obtained from solar spectral irradiance data collected by the Solar Radiation and Climate Experiment (SORCE)\footnote{\href{https://lasp.colorado.edu/home/sorce/data/ssi-data/}{lasp.colorado.edu/home/sorce/data/ssi-data}} \citep{solstice}.  


Using Equation \ref{ab_mag}, the magnitude of the Sun in the \textit{Kepler} passband was computed to be $-26.86$.
By scaling the flux density by a factor of $k\approx2.8646\times10^{-16}$, Equation \ref{ab_mag} will produce a magnitude of 12. 
So, to obtain the associated flux of a 12th magnitude solar type star, the total flux of the scaled spectral flux density through the Kepler passband was computed using Equation \ref{totalflux}:

\begin{equation}
	\label{totalflux}
	F = k\int f_\nu A(\nu) d\nu
\end{equation}


This computation suggests that a solar-type star with a Kepler magnitude of 12 produces a flux of $1.155\times10^{-13}\text{ W m}^{-2}$. Finally, using the pre-flight estimate from \citet{farmer_true_2013}, a conversion factor (Equation \ref{conversion_factor}) was derived to relate flux in electrons per second to flux in Watts per square metre for stars observed in the \textit{Kepler} passband:
\begin{equation}
1\text{ }\mathrm{e^-}/\text{s}=6.640\times10^{-19}\text{ W m}^{-2}
\label{conversion_factor}
\end{equation}

In our models, luminosities are computed in absolute units based on the mesh intensities, the distance, and third light of our systems. We can therefore constrain the passband luminosity of our models by sampling over temperature, radius, distance, gravity brightening and limb darkening coefficients, and generating all observables in the Kepler passband.

In PHOEBE, third light produces an additive effect on the light curve, which manifests as a vertical shift upwards in flux. Upon inspection of the Kepler Full Frame Images, the stars were found to be widely separated from neighboring stars, and there was no indication that any specific background sources needed to be considered. Since we could not obtain any prior information to constrain third light, we chose a value of 0 for the third light in each of our binary systems. If this assumption is incorrect, it would lead to an overestimate of the effective temperature and radius of each star.

\subsection{Polynomial Fitting to Detrend Kepler Data} \label{subsec:splinefit}

Data from the \textit{Kepler} spacecraft often exhibit significant systematic errors resulting from temperature gradients across the spacecraft, incident cosmic rays, and the planned quarterly rotation of the instrument \citep{stumpe_kepler_2012}. Such errors can obscure the intrinsic variability of stars and introduce artificial trends which complicate any comparisons between the data and synthetic models.

\begin{figure*}[!tb]
    \centering
    \includegraphics[width=1.0\textwidth]{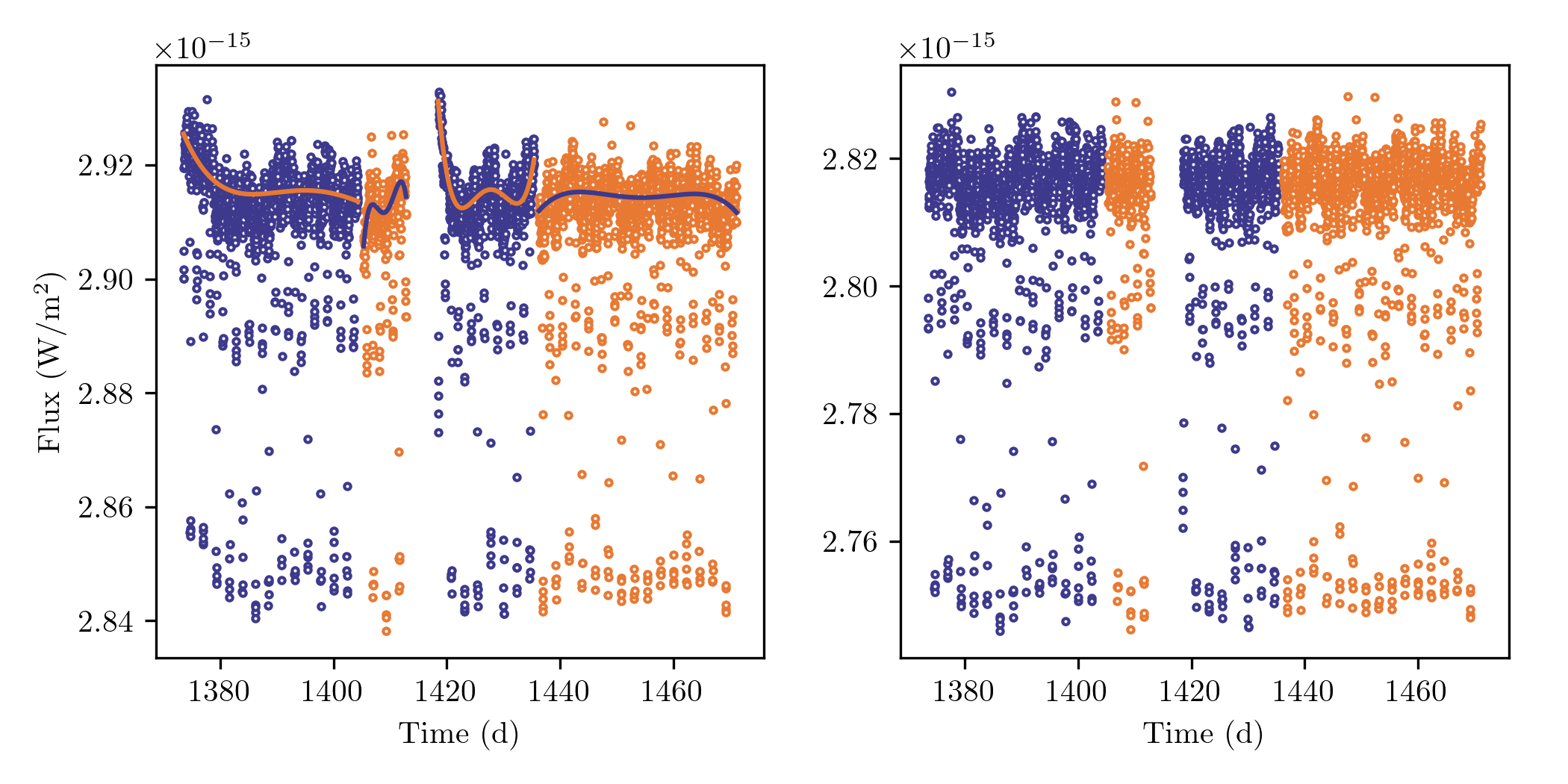}
    \caption{Left: fifth-order Legendre polynomials fitted to the second quarter of KIC 10727668 raw light curve data. Right: the final processed data obtained by subtracting off these polynomials and averaging the median flux of all quarters. Color alternates to indicate data gaps, where each subset of the data is fitted by a separate polynomial curve (solid line).}
    \label{fig:detrending}
\end{figure*}

Although presearch data conditioning modules are in place within the \textit{Kepler} data pipeline to correct for these errors \citep{stumpe_kepler_2012}, significant trends and discontinuities persisted in the photometric data acquired from the Barbara A. Mikulski Archive for Space Telescopes (MAST). To reduce the significance of such effects, the data were fitted with a set of fifth-order Legendre polynomials connected at each data gap in the light curve, as described in Section 2.6 of \citet{maxted_tess_2020} and shown in Figure \ref{fig:detrending} for KIC 10727668. For KIC 5957123 and KIC 8314879, linear regressions better characterised the trends, and were used instead of the Legendre polynomials to treat the data. Once all trends were flattened, the data was scaled to the median of all quarters.

The flux calibration was applied before adjusting the data to match the median of all quarters. By subtracting the fifth-order Legendre polynomials or linear regressions, we preserve light trends intrinsic to the binary that inform our MCMC sampling, such as ellipsoidal variation, and assume that any additional light trends come from outside the system, namely from the aforementioned errors in the Kepler satellite.

\subsection{Interstellar Extinction} \label{subsec:extinction}

Updated foreground-reddening estimates based on work by \citet{schlafly_measuring_2011} were obtained from the NASA Infrared Science Archive (IRSA)\footnote{\href{https://irsa.ipac.caltech.edu/applications/DUST/}{irsa.ipac.caltech.edu}} and used to estimate line-of-sight interstellar extinction values for each of the three binary systems. All three binaries are in a nearby region of our galaxy, so a value of $R_V=3.1$ was assumed. Given $A_V$ and $R_V$, our modelling software (Section \ref{subsec:phoebe}) can process extinction on the fly when computing the forward model.

\subsection{The Gaia Mission} \label{subsec:gaia}

Parallax measurements ($\pi''$) from the third Gaia data release \citep{brown_gaia_2020} were used to further constrain the binary models. Parallax values for these three star systems are detailed in Table \ref{tab:gaia}. These distance values, alongside the calibrated fluxes described in Section \ref{subsec:calibration}, constrain the passband luminosity of each binary.

\begin{table}[h!tb]
	\caption{Gaia Parallax Measurements}
	\label{tab:gaia}
		\begin{tabular}{ccc}
			\hline
			\hline
			Star & $\pi''$ & $d$ (pc)\\
			\hline
			KIC 5957123 & $0.27\pm0.02$ & $3700\pm300$\\
			KIC 8314879 & $0.13\pm0.02$ & $7700\pm1400$\\
			KIC 10727668 & $0.18\pm0.03$ & $5600\pm990$\\
			\hline
		\end{tabular}
	\label{gaia_data}
\end{table}

\section{Models} \label{sec:models}

A model was created for each of the three binary systems from which synthetic observables could be computed and compared to the observed photometry. Initial parameter fits were established as described in Section \ref{subsec:initialmodels}. Then, Markov Chain Monte Carlo methods were used to explore the parameter space and uncover posterior distributions for each system.

\subsection{PHOEBE} \label{subsec:phoebe}

The models analysed in this research were created using the PHysics Of Eclipsing BinariEs (PHOEBE)\footnote{\href{http://phoebe-project.org/}{phoebe-project.org}} stellar modelling package. Originally an adaptation of the 1971 Wilson-Devinney binary modelling code \textbf{\citep{wilson_realization_1971, prsa_computational_2005}}, PHOEBE has since been developed into a standalone binary analysis package which supports a number of inverse problem-solving techniques and data analysis tools \textbf{\citep{prsa_physics_2016, conroy_physics_2020}}. This research was performed using PHOEBE version 2.3.41. To aid in reproducibility, we will strive to record the specific settings used for all models and analysis tools throughout this work. For more comprehensive details, a Githib repository containing all PHOEBE models generated in this research has been released alongside this paper.\footnote{\href{https://github.com/podesse/computational-models-of-three-kepler-binaries-repo}{https://github.com/podesse/three-kepler-binaries-repo}}

For each PHOEBE model, synthetic light curves were computed through the \texttt{Kepler:mean} passband, and the passband luminosity mode was set to \texttt{absolute}. Distance estimates from Gaia and extinction estimates from the IRSA were then provided to complete the flux calibration. In most systems, gravity darkening and reflection coefficients were set to 0.9 and 1.0 respectively, as these values are characteristic of stars with radiative atmospheres. However, both these coefficients were sampled in the later MCMC analysis, as it was not always clear which specific value was acceptable given the temperature of the star. This is especially prevalent when modeling stars in the 6000K to 8000K temperature range, for which it is not always obvious if a convective atmosphere or radiative atmosphere is more appropriate.

Synthetic observables were generally computed at 100 time steps over one full orbit. 
This synthetic light curve was then extrapolated to the time series of the full data set. The eclipses of KIC 10727668 were more densely sampled, so this system's forward model was computed at a total of 119 time steps to better resolve the model during eclipses.

Limb darkening lookup tables were used to interpolate the limb darkening coefficients for most stars with PHOEBE's logarithmic interpolation law. Generally, \texttt{ck2004} model atmospheres \citep{castelli_new_2004} were used when determining the emergent passband intensity of most stars. The only exception was the primary star in the model of KIC 10727668, which appears to be a White Dwarf star. For this star, we supplied limb darkening coefficients manually from \citet{claret_gravity_2020}, and used a blackbody atmosphere model rather than \texttt{ck2004} model atmospheres, as support for White Dwarf atmospheres in PHOEBE is still under development.

For full data set analysis, the noise-nuisance parameter $\sigma_{\text{lnf}}$ was introduced, which is detailed further in section \ref{subsec:likelihood-function}. Most MCMC runs suggested a value for $\sigma_{\text{lnf}}$ between $-6$ and $-7$. The most reliable method of identifying an appropriate range for $\sigma_{\text{lnf}}$ seems to be through repeated tests of different values.

\subsection{Procedure for Analysis} \label{subsec:procedure}

First, solvers bundled with PHOEBE were used to establish some fundamental parameters of the binaries. The light curve periodogram was applied to the full data set to estimate the orbital period. Then, the light curve geometry solver was used to establish the time of superior conjunction ($t_{\text{0,supconj}}$), the orbital eccentricity ($e$), and the argument of periastron ($\omega$) of the system. It should be noted that this estimator only works with eclipsing binaries, so these values were fit manually for the ellipsoidal variable system KIC 5957123.

Any non-zero value for the orbital eccentricity greatly increases computation time, but in many cases its effect on the light curve geometry could not be ignored. In further analysis, the parameters $e\sin\omega$ and $e\cos\omega$ were sampled over rather than the eccentricity and the argument of periastron, since $e$ and $\omega$ are highly correlated whereas $e\sin\omega$ and $e\cos\omega$ are more orthogonal.

Next, parameter values were manipulated manually to obtain an initial fit, followed by use of PHOEBE's Nelder-Mead optimizer to improve on these manual estimates. This greatly helps reduce the burn-in time of the MCMC sampler, and allowed us to define more reasonable initializing distributions.

Finally, MCMC sampling via the emcee python package \citep{foreman-mackey_emcee_2013} was used to obtain estimates of the posterior distributions on all of our parameters. Initially, light curve data was phase-folded and averaged into 500 bins, and uncertainty on each bin was assigned based on the standard deviation of points in that bin. An initial MCMC run was performed on this binned light curve data to encourage fast convergence and obtain information on the expected shape of the posterior distribution. Once adequate sampling in the binned data set run was complete, the resulting posterior distributions were used as initializing distributions for MCMC analysis of the full data set, from which more robust posteriors may be derived. This follow-up run was necessary as analysis of a binned light curve generally does not yield appropriate uncertainty estimates. This issue is discussed in more detail in Section \ref{subsec:posterior} and Section \ref{subsec:parameter_uncertainties}.


\subsection{Initial Parameters} \label{subsec:initialmodels} 

\begin{figure*}[tb]
    \centering
    \includegraphics[width=1.0\textwidth]{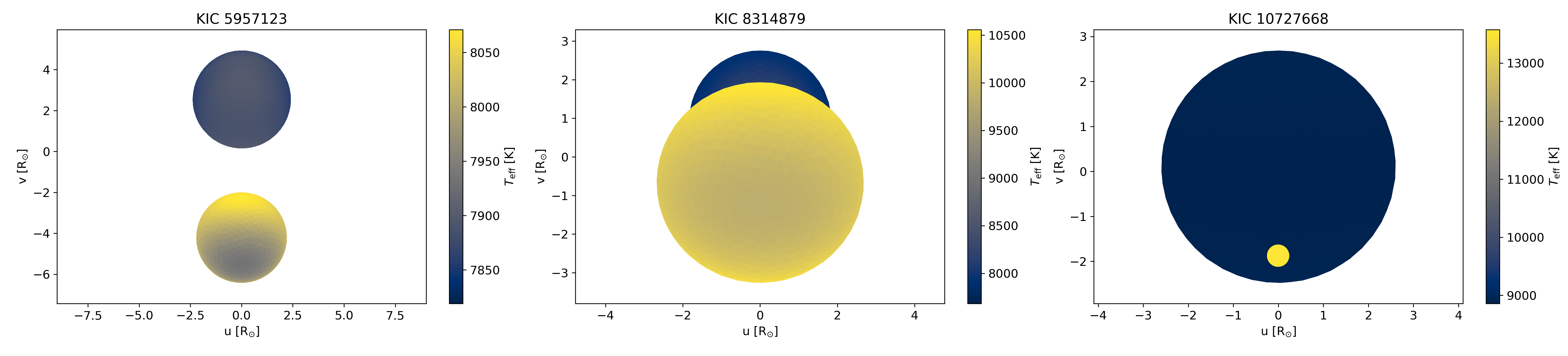}
    \caption{Mesh plots illustrating the geometry of each system. All plots are displayed at phase $\phi=0.5$, corresponding to the time of secondary eclipse when the primary star eclipses the secondary. KIC 5957123 does not eclipse, KIC 8314879 eclipses only partially, and the White Dwarf primary of KIC 10727668 fully transits its companion.}
    \label{fig:mesh_all}
\end{figure*}

Initial solutions were obtained for each system using  PHOEBE solvers, manual fitting, and pre-published estimates (where available). We cannot generalize the procedure used to obtain initial solutions, as each system required unique treatment. Instead, we will detail how we arrived at an initial model for each system. Initial parameters for each model are displayed in Table \ref{tab:initial_parameters}, and the geometry of each system at the time of secondary eclipse is shown in the mesh plots of Figure \ref{fig:mesh_all}.

\begin{table*}[h!tb]
\begin{center}
	\caption{Initial Model Parameters} 
		\begin{tabular}{ccccc}
	    	\hline
			\hline
			Parameters & Symbol & KIC 5957123 & KIC 8314879 & KIC 10727668 \\
			\hline
			\multicolumn{2}{c}{\textit{System Parameters}}\\ 
			Period & $P$ & 2.82232 d & 0.87776 d & 2.30592 d\\
			Superior Conjunction & $t_{\text{{0,supconj}}}$ & 1274.621 d & 1274.578 d & 1275.519 d\\
			Inclination & $i$ & 54.99$^\circ$ & 77.99$^\circ$ & 79.42$^\circ$\\
			Mass Ratio & $q$ & 1.650 & 0.6916 & 17.31\\
			Semi-Major Axis & $a$ & 11.89$R_\odot$ & 7.694 $R_\odot$ & 10.67 $R_\odot$\\
			Temperature Ratio & $\frac{T_{\text{eff,2}}}{T_{\text{eff,1}}}$ & 0.9814 & 0.7662 & 0.6559\\
			Radius Ratio & $\frac{R_2}{R_1}$ & 1.082 & 0.6838 & 10.96\\
			Fractional Radii & $\frac{R_1 + R_2}{a}$ & 0.3837 & 0.5847 & 0.2640\\
			Distance & $d$ & 3715 pc & 7902 pc & 5325 pc\\
			Eccentricity & $e$ & 0.01037 & 0.0 & 0.01038\\
			Arg. of Periastron & $\omega$ & 271.3$^\circ$ & 0.0$^\circ$ & 88.86$^\circ$\\
			\hline
			\multicolumn{2}{c}{\textit{Stellar Parameters -- Primary}}\\ 
			Effective Temperature & $T_{\text{eff,1}}$ & 8005 K & 10194 K & 13546 K\\
			Equivalent Radius & $R_{\text{equiv,1}}$ & 2.190$R_\odot$ & 2.672$R_\odot$ & 0.2354$R_\odot$\\
			Mass & $M_1$ & 1.067$M_\odot$ & 4.689$M_\odot$ & 0.1673$M_\odot$\\
			Gravity Brightening & Gr. Bright. & 0.7578 & 0.9477 & 0.9463 \\
			Reflection Fraction & Refl. Frac. & 0.7479 & 0.9462 & 0.9006 \\
			\hline
			\multicolumn{2}{c}{\textit{Stellar Parameters -- Secondary}}\\ 
			Effective Temperature & $T_{\text{eff,2}}$ & 7856 K & 7811 K & 8885 K\\
			Equivalent Radius & $R_{\text{equiv,2}}$ & 2.370$R_\odot$ & 1.827$R_\odot$ & 2.581$R_\odot$\\
			Mass & $M_2$ & 1.761$M_\odot$ & 3.243$M_\odot$ & 2.897$M_\odot$\\
			Gravity Brightening & Gr. Bright. & 0.7621 & 0.7559 & 0.9509\\
			Reflection Fraction & Refl. Frac. & 0.7645 & 0.8646 & 0.9037\\
			\hline
		\end{tabular}
	    \label{tab:initial_parameters}
\end{center}
\end{table*}

\subsubsection{KIC 5957123}

The light curve geometry of this system presented several difficulties when trying to identify an initial solution. KIC 5957123 appears to be an ellipsoidal variable system with a period $P\approx2.8223$ days. Often, initial solutions identified a small semi-major axis, which caused the model to adopt low masses uncharacteristic of stars at this temperature. A more robust initial solution was eventually obtained by re-arranging several constraints to manipulate sums and ratios of parameters, and manually fitting to reasonable values.



Rather than manipulate temperatures and radii directly, the sums and ratios $\frac{T_{\text{eff,2}}}{T_{\text{eff,1}}}$, $\frac{R_{2}}{R_{\text{1}}}$, and $\frac{R_{1}+R_{2}}{a}$, alongside $T_{\text{eff,1}}$, were sampled over when solving for the initial solution. First, the semi-major axis $a$ was scaled to a value which yielded masses greater than $1M_\odot$, as temperature estimates from the Kepler Input Catalogue (KIC) suggested $T_{\text{eff}} > T_{\text{eff,}\odot}$. To manipulate the fractional equivalent radii $\frac{R_{1}+R_{2}}{a}$ to fit the synthetic light curve amplitude to the photometric data, the passband luminosity mode ({\texttt{pblum\_mode}}) was temporarily switched to \texttt{dataset-scaled}. In this mode, PHOEBE automatically adjusts the passband luminosity so that the synthetic light curve is calibrated to the observed light curve. Generally, this behaviour is undesirable as it prevents us from properly constraining the luminosity of the system. In this case, however, it allowed us to fit $\frac{R_{1}+R_{2}}{a}$ without requiring additional calibration of the distance or primary effective temperature, both of which scale the light curve without significantly altering its shape. 
Then, holding $\frac{R_{1}+R_{2}}{a}$ constant, {\texttt{pblum\_mode}} was switched back to {\texttt{absolute}} whereby the luminosity of the system is fully determined by the intensity of the stellar surfaces and the distance, and the light curve was scaled to match the observed data by manipulating $T_{\text{eff,1}}$ and $a$. Finally, values for $\frac{T_{\text{eff,2}}}{T_{\text{eff,1}}}$, $\frac{R_{2}}{R_{1}}$, $e\cos\omega$, and $e\sin\omega$ were adjusted as appropriate to obtain an adequate initial fit. These values are displayed in Table \ref{tab:initial_parameters}.

\subsubsection{KIC 8314879}

The light curve for KIC 8314879 has pointed eclipses indicative of a partially eclipsing binary. Some ellipsoidal variation is apparent between eclipses, suggesting that these two stars orbit in close proximity. Initial parameter estimates were quickly derived for this system using PHOEBE's estimators and some manual fitting, followed by optimization with the Nelder-Mead optimizer. These parameter estimates are summarized in Table \ref{tab:initial_parameters}.

\subsubsection{KIC 10727668}

In an eclipsing binary, the deeper eclipse occurs when the hotter star is obscured. A flat-bottomed eclipse occurs if the smaller of the two stars is completely obscured. The deeper eclipse of KIC 10727668's light curve has a flat eclipse bottom, a combination indicative of a White Dwarf component with a F-, A-, or B-type companion \citep{rappaport_discovery_2015}. Previous investigations of this system support this assumption \citep{rappaport_discovery_2015, wang_v1224_2018}.

Taking the primary star to be the star at superior conjunction near orbital phase $\phi=0$ \citep{phoebe_scientific_ref}, 
we assigned parameters characteristic of a White Dwarf to the primary component. These parameters are presented in Table \ref{tab:initial_parameters}. Support for White Dwarf atmospheres in PHOEBE is still under development, so our primary star was computed with a blackbody atmosphere. Limb-darkening coefficients were assigned manually for this star; White Dwarf limb-darkening coefficients were obtained from \citet{claret_gravity_2020}. Based on the temperature and surface gravity ($T_{\text{eff,1}}=13526\text{K}$, $\log g_1 = 4.95$) of the White Dwarf in our initial model, we chose bolometric limb darkening coefficients of [0.4885 0.3903].\footnote{These specific coefficients can be found on line 21860 of tableef.dat from \citet{claret_gravity_2020}.}These are the logarithmic limb darkening coefficients $e$ and $f$ respectively for the Kepler passband.

\section{Markov chain Monte Carlo} \label{sec:mcmc}

Generally, both radial velocity data and photometric data are required to constrain the full physical properties of an eclipsing binary, with radial velocities constraining stellar masses, radii, effective temperatures, and the orbital semi-major axis, as detailed in Table 1 of \citet{southworth_binary_2020}. As there remain several degeneracies we cannot eliminate without spectroscopic data, any solution we propose is bound to have a high degree of uncertainty. To account for this degeneracy, we include these unconstrained parameters in our Markov Chain Monte Carlo (MCMC) analysis. By drawing upon unconstrained parameters, we obtain a more realistic sampling of the parameter space.



We initialized our MCMC sampler with 64 walkers, a value chosen to simplify parallelization across 64 CPUs. These walkers were initialized at positions drawn from a Gaussian distribution around our initial parameter estimates, the width of which was tuned based on the sensitivity of each system's light curve to different parameter values. To achieve fair and adequate sampling, the ideal number of iterations is over 100 times the auto-correlation time for each parameter. Given the auto-correlation times characteristic of our models, this would mean processing between 10,000 and 100,000 iterations.

Generally, we sampled up to 15 parameters as listed in Table \ref{tab:initial-distribution}. The widths presented for Gaussian distribution types correspond to $1\sigma$ widths. Meanwhile, for uniform distributions, the range of values covered by that distribution is displayed.
\begin{table}[h!tb]
    \centering
    \caption{Example Initializing Distribution}
    \begin{tabular}{ccc} 
    \hline
    \hline
        Parameter & Width & Distribution \\
        \hline
        $i$ &  $2.0^\circ$ & Gaussian \\
        $q$ & $0.1$ & Gaussian \\
        $a$ & $0.1 R_\odot$ & Gaussian \\
        $T_{\text{eff,1}}$ &  $100$ K & Gaussian\\
        $\frac{T_2}{T_1}$ & 0.05 & Gaussian \\
        $\frac{R_2}{R_1}$ & 0.05 & Gaussian \\
        $\frac{R_1+R_2}{a}$ & 0.025 & Gaussian \\
        $e\sin\omega$ & 0.01 & Gaussian \\
        $e\cos\omega$ & 0.01 & Gaussian \\
        $d$ & $160$ pc & Gaussian\\
        Gr. Bright.$_{1,2}$& [0.9, 1.0] & Uniform\\
        Refl. Frac.$_{1,2}$ & [0.8, 1.0] & Uniform \\
        $\sigma_{\text{lnf}}$ & 1 & Gaussian\\
        \hline
    \end{tabular}
    \label{tab:initial-distribution}
\end{table}

In the analysis of KIC 5957123, the time of superior conjunction, $t_{\text{0,supconj}}$, was also sampled, as the light curve geometry estimator could not be applied to obtain an adequate fit for $t_{\text{0,supconj}}$ prior to this analysis. Initial estimates for the eccentricity of KIC 8314879 found $e=0.088$, which is very close to circular. Thus, $e\cos\omega$ and $e\sin\omega$ were not sampled for this system, since it was found that models with strictly circular orbits ($e=0$) could be computed nearly 4 times as quickly than those with an eccentric orbit. However, more realistic uncertainties may be achieved by sampling $e\cos\omega$ and $e\sin\omega$. 

When sampling a full data set, we included a noise-nuisance parameter, $\sigma_{\text{lnf}}$, to account for observational noise and underestimated per-point uncertainties. This parameter is described in more detail in Section \ref{subsec:likelihood-function}. 
We chose to sample over the sums and ratios ($\frac{T_{\text{eff,2}}}{T_{\text{eff,2}}}$, $\frac{R_2}{R_1}$, and $\frac{R_1+R_2}{a}$, rather than absolute parameter values ($T_{\text{eff,2}}$, $R_1$, $R_2$) because these sums and ratios are more directly constrained by the photometric data (see Figure 11 of \citet{conroy_physics_2020} for a similar comparison of sampling orthogonal parameters). These parameter combinations produce posteriors which are less correlated and closer to Gaussian distributions, enabling a more robust interpretation of parameter values. 

MCMC methods sample the posterior distribution, which describes the probability that a given parameter combination generates our data. This concept follows from Baysean statistics: Bayes' Theorem is listed in Equation \ref{bayesthm}, which gives the posterior probability $\mathcal{P}(\Theta)$  of a parameter combination $\Theta$ \citep{speagle_conceptual_2020}:
\begin{equation}
    \label{bayesthm}
    \mathcal{P}(\Theta)= \frac{\mathcal{L}(\Theta)\pi(\Theta)}{\mathcal{Z}}
\end{equation}
where $\pi(\Theta)$ is the prior probability (see section \ref{subsec:prior-dist}) and $\mathcal{L}(\Theta)$ is the likelihood function (see section \ref{subsec:likelihood-function}). The evidence, $\mathcal{Z}$, is difficult to compute, so we ignore the absolute value of our distributions and perform the simplification shown in Equation \ref{simplebayes} \citep{hogg_data_2010, prsa_mcmc_2021}:
\begin{equation}
    \label{simplebayes}
    \mathcal{P}(\Theta)\propto\mathcal{L}(\Theta)\pi(\Theta)
\end{equation}
Inherent in this simplification is the assumption that our model appropriately represents the data we are given. We will cover posterior distributions in more detail in section \ref{subsec:posterior}, but for now it should be noted that the width of the posterior distributions are not necessarily indicative of parameter uncertainties. 

\subsection{Prior Distributions}
\label{subsec:prior-dist}

For most parameters sampled by MCMC, we did not introduce custom prior distributions, since prior information was only available for a select few parameters. PHOEBE intrinsically places some restrictions on the forward model, and runs checks to ensure that the model is physical before computing. These restrictions act as uninformative priors in extreme cases where walkers attempt to sample regions of the parameter space which are obviously nonphysical. 


A Gaussian prior was introduced over the distance for each binary system, with standard deviation equal to the measurement uncertainties published by Gaia, as shown in Table \ref{gaia_data}.

Uniform (uninformative) priors were placed on gravity darkening and reflection coefficients based on the temperatures of the primary and secondary stars from the initial solutions. For stars with $T_{\text{eff}}>8000\text{ K}$, we expect radiative atmospheres for which the most appropriate gravity darkening and reflection coefficients are $\geq 0.9$ and $\geq 0.8$ respectively. For stars with high temperature initial solutions, uniform priors were set over this range. There are a few systems where the effective temperature of at least one star is within the range of $6000\text{ K} \leq T_{\text{eff}} \leq 8000 \text{ K}$. For such stars, the appropriate gravity darkening and reflection coefficients are somewhat ambiguous; with the models available to us in PHOEBE, we cannot determine whether the stellar atmosphere should be radiative or convective for a given set of parameters. Therefore, for stars with $ 6000\text{ K}\leq T_{\text{eff}}\leq8000\text{ K}$, uniform priors were set which enclose the entire range of possible values, including gravity darkening and reflection coefficients in the range of $[0.32, 1.0]$ and $[0.6, 1.0]$ respectively.

\subsection{The Likelihood Function}
\label{subsec:likelihood-function}

To each parameter combination, the MCMC method assigns a probability based on the chosen likelihood function. With PHOEBE, this involves computing the forward model, computing residuals between the model and the observed data, and performing a $\chi^2$ evaluation. In this research, we used \texttt{PHOEBE}'s default log-likelihood function, the exact form of which is given in Equation \ref{loglikelihood}. Here, the log of probability, $p$, is given as a sum in terms of the model data $y_m$, the observed data $y_o$, the per-point uncertainty on the observed data, $\sigma_o$, and our noise-nuisance parameter $\sigma_{\text{lnf}}$:
\begin{equation}
    \label{loglikelihood}
    \ln(p) = -0.5\sum\bigg(\frac{(y_m-y_o)^2}{\sigma^2}+\ln(\sigma^2)\bigg)
\end{equation}
where
\begin{equation}
    \label{sigma_denom}
    \sigma^2 = {\sigma_o}^2+{y_m}^2e^{2\sigma_{\text{lnf}}}
\end{equation}
It is worth bringing to attention the ways in which Equation \ref{loglikelihood} differs from the expected, simpler form of the log-likelihood function, $\ln(p)=-0.5\chi^2$, notably through the addition of a $\ln(\sigma^2)$ term in the sum. The denominator (Equation \ref{sigma_denom}) contains an additional term ${y_m}^2e^{2\sigma_{\text{lnf}}}$. The convergence of an MCMC run and the resulting widths of the posterior distributions are both highly dependent on the per-point observational uncertainties \citep{conroy_physics_2020}. In their paper on solving the inverse problem with PHOEBE, \citet{conroy_physics_2020} explain that in cases where the observational uncertainties are thought to be underestimated by a constant factor (such as in \textit{Kepler} data; see \citet{jenkins_kepler_2017}) the noise-nuisance parameter is introduced to increase the scale of the observational uncertainties. We sample over $\sigma_{\text{lnf}}$ in our MCMC analysis so that any resulting degeneracies with other nuisance parameters are encoded in the posteriors.

The additional $\ln(\sigma^2)$ term serves as somewhat of a counter to the noise-nuisance parameter. If we do not wish to include the noise-nuisance parameter in our computation, we assign $\sigma_{\text{lnf}}$ a value of $-\infty$; this simplifies our uncertainty term to a constant $\sigma^2={\sigma_o}^2$. In such a case, we are effectively adding an arbitrary constant to all of our log-likelihood calculations, which does not impact the MCMC walkers' behaviour in any significant way.

If instead we elect to include the noise-nuisance parameter (taking $\sigma_{\text{lnf}}$ to be some finite value), we expect Equation \ref{sigma_denom} to increase, leading to a more favorable log-likelihood estimate. The $\ln(\sigma^2)$ term in Equation \ref{loglikelihood} prevents $\ln(p)$ from improving arbitrarily as larger values of $\sigma_{\text{lnf}}$ are sampled, since $\ln(\sigma^2)$ increases linearly with $\sigma_{\text{lnf}}$ whereas the $\chi^2$ term decreases exponentially with $\sigma_{\text{lnf}}$. Once the noise-nuisance parameter suitably represents the noise of the data, any further improvement we gain in $\chi^2$ by increasing $\sigma_{\text{lnf}}$ is smaller than the detriment incurred upon the log-probability by the $\ln(\sigma^2)$ term. Modifying the log-probability function in this manner deters overfitting data noise when sampling over $\sigma_{\text{lnf}}$.


\section{Results of MCMC Analysis} \label{sec:results}

We have performed MCMC analysis on both binned and full data sets for all three binary systems, uncovering the posterior distributions in the region of the parameter space near our initial solution. The topology of the posterior distribution, especially the modality, should indicate whether or not the MCMC runs are adequately converged \citep{hogg_data_2018}. Although MCMC sampling cannot guarantee that we have identified the global minimum of the parameter space, a converged MCMC run that is independent of initialization is likely to have enclosed the global minimum \citep{hogg_data_2018}. 

\subsection{Posterior Distributions}
\label{subsec:posterior}


Posterior distributions indicate the probability that a given parameter set will describe the observed data. The shapes of the posterior distributions provide us with some knowledge of the local topology of the parameter space near our solution, indicating parameter correlations and regions of confidence around our parameter estimates. 

When drawing parameter uncertainties directly from the widths of posterior distributions, however, we run the risk of misinterpreting the results of our MCMC sampling. We must ensure that we have marginalized over all nuisance parameters -- any parameter that has a significant effect on the shape of the light curve -- or else we may fail to sample all possible parameter combinations which yield an acceptable fit for our system.

We must also ensure that our MCMC walkers have completed an adequate number of iterations to both converge on the global solution and fully explore that region the parameter space. If too few iterations are completed, then the walkers may not sample an acceptable volume of the posterior distribution, leading to smaller than expected posterior widths.

Finally, we must ensure that our model adequately represents the data provided. As discussed in section \ref{subsec:likelihood-function}, \textit{Kepler} data contains stochastic noise and underestimated errors which our model attempts to fit with a noise-nuisance parameter. This method often struggles to converge, in which case we turn to performing MCMC analysis on binned light curve data. With larger per-point uncertainties and fewer data points, this binned data enables the MCMC walkers to converge quickly and more readily accept new positions, leading to a fuller exploration of the parameter space. However, binned data is not necessarily representative of the binary system observed by \textit{Kepler}, and leads to posterior distributions with larger widths than our observational precision would allow. Therefore, we must strive to strike a balance between data which encourages convergence and a model which appropriately generates the observed data.


Many parameters were sampled in the MCMC analysis of these systems. As such, the corner plots displaying 2-dimensional cross sections of the posterior distributions are quite large. To aid in legibility, full corner plots for each run are shown in Appendix \ref{appendix:cornerplots}. The plots displayed in the following subsections do not show the full array of sampled parameters, but rather highlight specific parameter correlations within the larger distribution collection.

\begin{figure}[!tb]
    \centering
    \includegraphics[width=0.45\textwidth]{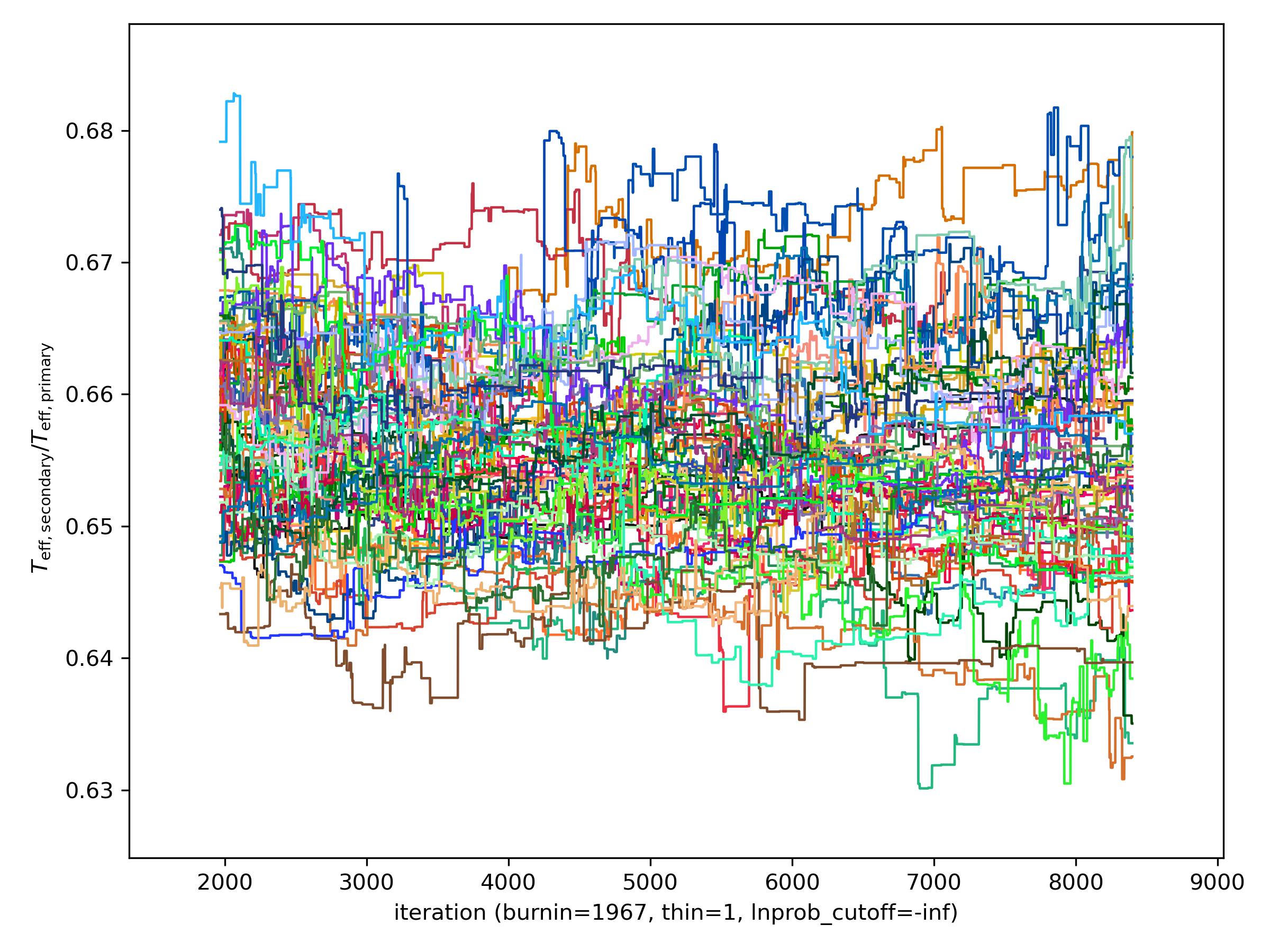}
    \caption{An example of the trace plot from the MCMC analysis of KIC 10727668 using binned data. Each chain is represented by a different colour, and every parameter value sampled is shown. This plot examines $\frac{T_2}{T_1}$, which has an estimated autocorrelation time of 896.8 iterations. The walkers readily take on new parameter combinations, and the ensemble of walkers appears to have traversed the parameter space multiple times.}
    \label{fig:trace_example}
\end{figure}

Although we cannot practically know whether the MCMC walkers have fully sampled the posterior distribution, we can define some heuristic characteristics, such as autocorrelation time and acceptance fraction, to evaluate the completion of these MCMC runs. 
\citet{hogg_data_2018} suggest that an MCMC run has completed adequate sampling when each walker has traversed the high-probability region of the parameter space multiple times. More precisely, \citet{hogg_data_2018} define the autocorrelation time per-parameter as the number of steps needed for the walkers to draw independent samples of that parameter. 
This can be examined qualitatively with a trace plot, like the one shown in Figure \ref{fig:trace_example}. The MCMC sampler is performing well if the chains overlap and take on parameter values across the entire range of values sampled. Ideally, we may consider an MCMC run as converged when the number of iterations completed is 10 to 100 times greater than the largest autocorrelation time. 

The acceptance fraction is defined as the ratio of moves accepted by a walker to the total number of iterations completed. In the extreme cases, an acceptance fraction of 0 indicates that the walkers never change position, whereas an acceptance fraction of 1 indicates that the walkers accept every new position proposed. An acceptance fraction in the range of 0.25 to 0.5 is most desirable, striking a balance between convergence and exploration of the parameter space \citep{hogg_data_2018}. We comment on how the acceptance fraction may be improved for our models in Section \ref{subsec:feasibility}.


\subsection{KIC 5957123}


\begin{figure*}
\centering
\begin{minipage}[b]{.45\textwidth}
    \centering
    \includegraphics[width=1.0\textwidth]{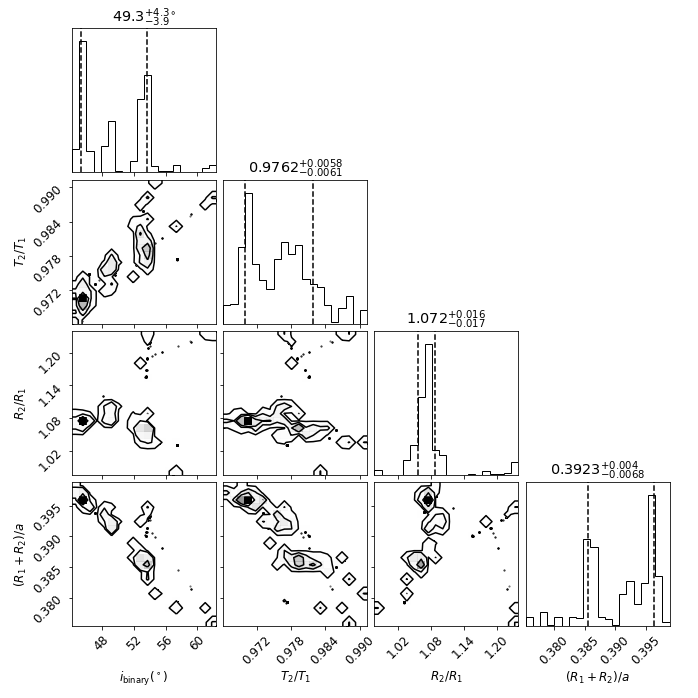}
\end{minipage}\qquad
\begin{minipage}[b]{.45\textwidth}
    \centering
    \includegraphics[width=1.0\textwidth]{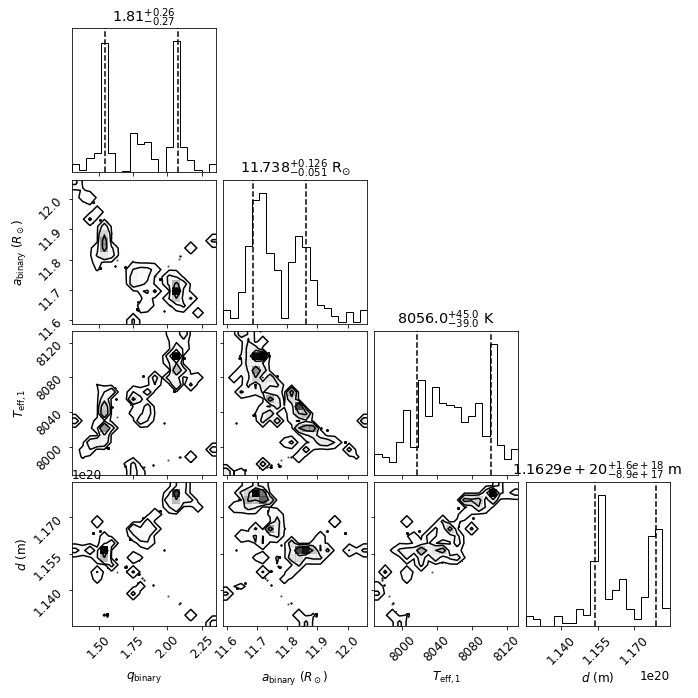}
\end{minipage}
\caption{Subsets of the corner plot of the posterior distributions uncovered by MCMC analysis of KIC 5957123. The left corner plot contains parameters which we expect to be constrained by photometric data ($i$, $\frac{T_2}{T_1}$, $\frac{R_2}{R_1}$, $\frac{R_1+R_2}{a}$), while the right plot shows a collection of parameters sampled as nuisance parameters ($q$, $a$, $T_{\text{eff,1}}$, $d$). One-dimensional posterior distributions for each parameter are displayed along the main diagonal, where the dashed lines indicate the $1\sigma$ spread of samples.}
\label{fig:cornerplots-5957123}
\end{figure*}

\begin{figure}[!t]
    \centering
    \includegraphics[width=0.45\textwidth]{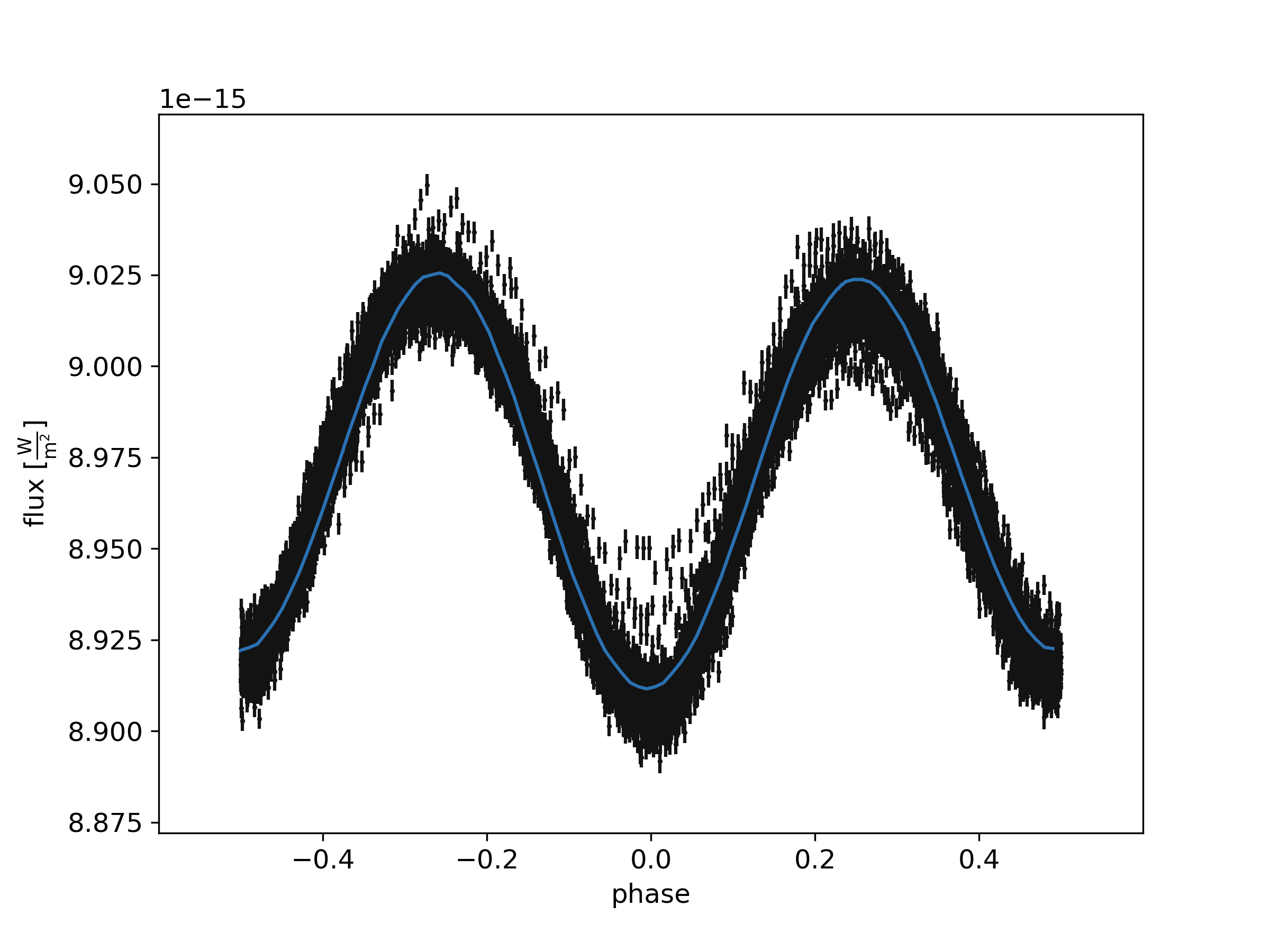}
    \includegraphics[width=0.45\textwidth]{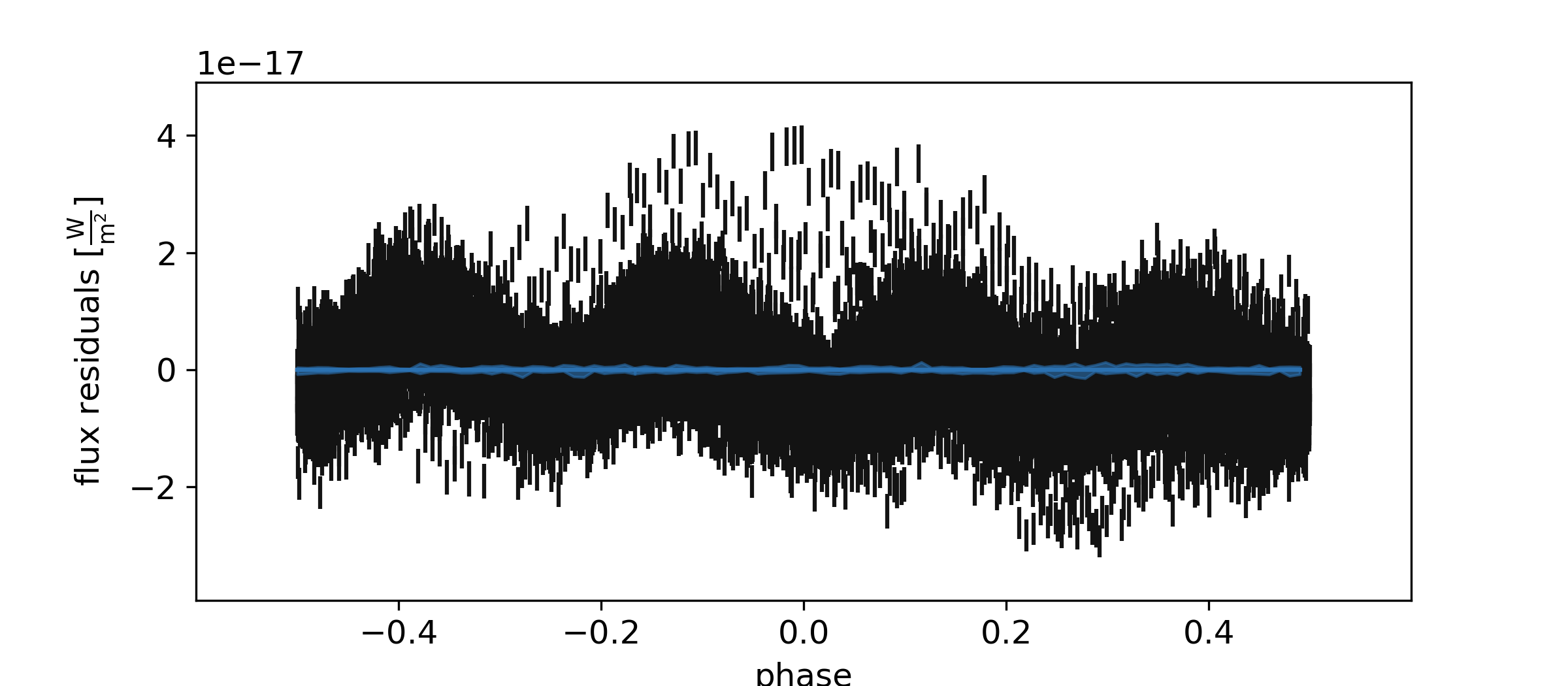}
    \caption{Phase-folded light curve of KIC 5957123 overplotted with 10 synthetic solutions drawn from the posterior distributions. Below, residuals between these synthetic models and the photometric data are plotted. These residuals exhibit phase-space trends, suggesting that our model does not generate observables that agree with the photometric data.}
    \label{fig:lc_5957123}
\end{figure}

A total of 7600 iterations were completed in the MCMC analysis of KIC 5957123. Autocorrelation times for each parameter were estimated using emcee's autocorrelation function, given by Equation 12 of \citet{foreman-mackey_emcee_2013}. The average autocorrelation time for the parameters was estimated to be $483\pm117$ iterations, however autocorrelation times could not be estimated for the inclination ($i$) or the secondary gravity brightening coefficient. This suggests that too few iterations have been completed to estimate the autocorrelation time, or that sampling of these two values is insufficient due to a poor initializing distribution. However, we suspect that our model may not appropriately generate the data, in which case fair sampling of the parameter space will never be achieved. This concern is explored in more detail throughout this section. The average acceptance fraction for this MCMC run was 0.0001. Notably, one walker had an acceptance fraction of 0, suggesting that it did not once change position throughout the 7600 iterations of sampling.


Since the light curve for KIC 5957123, shown in Figure \ref{fig:lc_5957123}, exhibits no distinct eclipses, the effects of mass ratio, temperatures, and stellar radii cannot be decoupled. We therefore expect to uncover broader and less centralized posteriors in our MCMC sampling with strong correlations between parameters.  
The posterior for $q$, seen in the right corner plot of Figure \ref{fig:cornerplots-5957123}, is distinctly multi-modal, and posteriors for $i$, $d$, $a$, and $\frac{R_1+R_2}{a}$ appear to have smaller secondary peaks. These posteriors suggest that the MCMC walkers have not converged on a single solution, and as a consequence, we cannot infer parameter uncertainties from their widths. Furthermore, the contour plots in both Figure \ref{fig:cornerplots-5957123} and the full corner plot shown in Figure \ref{fig:corner_full_5957123} suggest correlations between several parameters. Due to the sparse sampling of parameter combinations, these correlations are difficult to quantify, however in many of the two-dimensional histograms such as those relating $i$ and $\frac{T_2}{T_1}$ in Figure \ref{fig:cornerplots-5957123}, samples appear to lie along a line rather than clustered around one central point. Strong correlations were not present between these parameters in the binned data set analysis (see Figure \ref{fig:corner_binned_5957123}, where the posteriors are more centralized), so the correlations seen in the full data set posteriors may be a result of walkers stuck in separate regions of the parameter space. Nonetheless, this indicates that our model is struggling to generate the observed data.

The residuals in Figure \ref{fig:lc_5957123} show phase-space periodic trends, indicating that our synthetic light curve is a poor fit to the observed data. Additionally, the residuals at phase $-0.4$ are more positive than those at phase $0.4$, suggesting that one peak of the light curve is higher than the other. Unfortunately, our model of KIC 5957123 does not account for this phenomenon. Although we marginalized over many nuisance parameters, the failure of this MCMC run to converge indicates that this difference in peak brightness must result from some additional parameters that we have not considered. Therefore, our model cannot appropriately generate the data, and as such the MCMC walkers will never converge. Realizing this, we stopped this run early at 7600 iterations.

\subsection{KIC 8314879}


\begin{figure*}
\centering
\begin{minipage}[b]{.45\textwidth}
    \centering
    \includegraphics[width=1.0\textwidth]{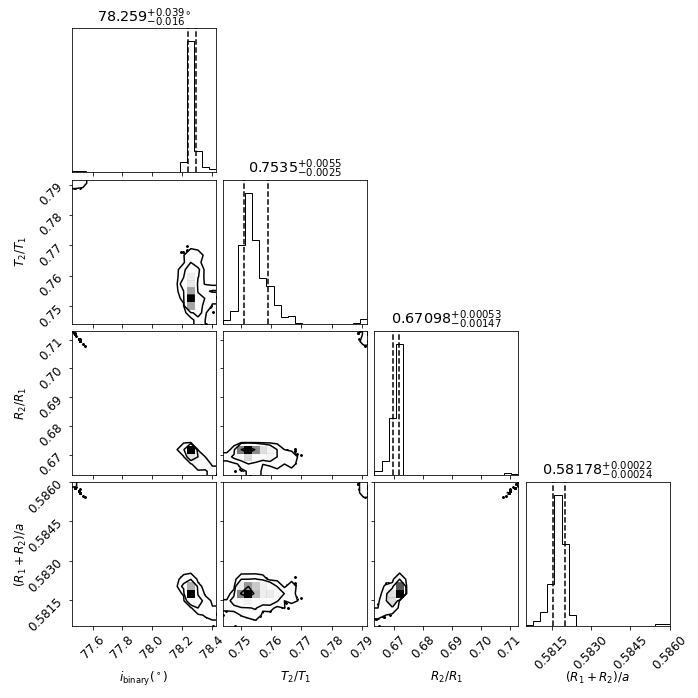}
\end{minipage}\qquad
\begin{minipage}[b]{.45\textwidth}
    \centering
    \includegraphics[width=1.0\textwidth]{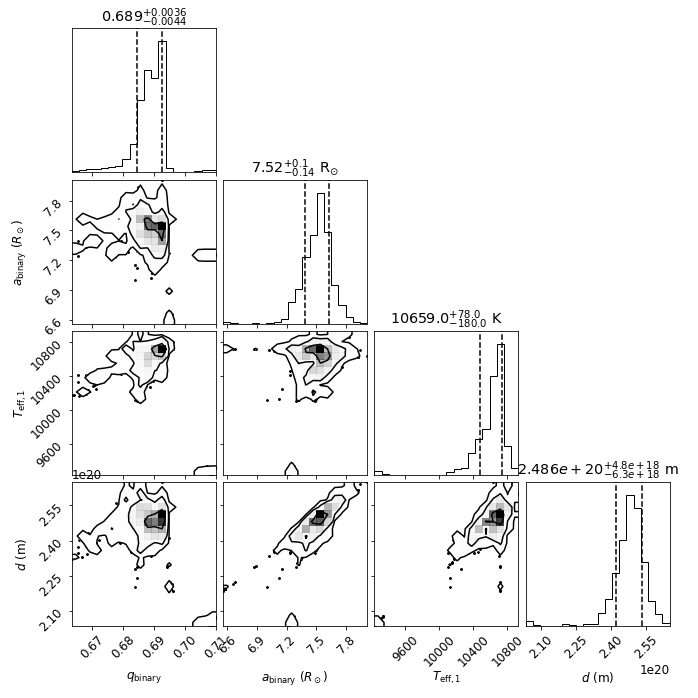}
\end{minipage}
\caption{Similar to Figure \ref{fig:cornerplots-5957123}, but for the posteriors uncovered by MCMC analysis of KIC 8314879. The log-probability was cut off at 516555, eliminating 1 walker from this solution. Posterior distributions for these parameters are predominantly uni-modal, although three walkers found a separate high-probability region visible along the edges of several posteriors.}\label{fig:cornerplots-8314879} 
\end{figure*}

\begin{figure}[!t]
    \centering
    \includegraphics[width=0.45\textwidth]{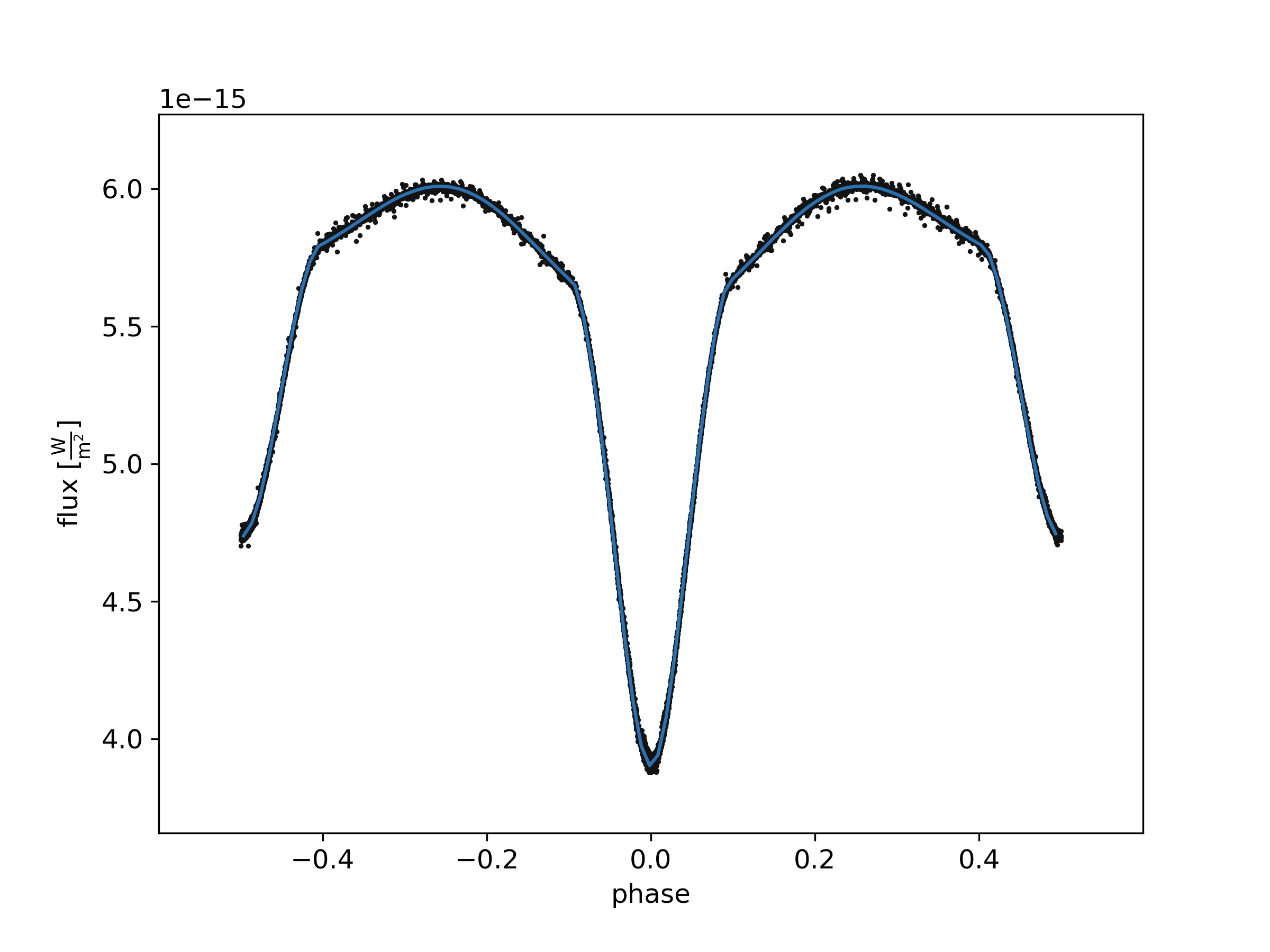}
    \includegraphics[width=0.45\textwidth]{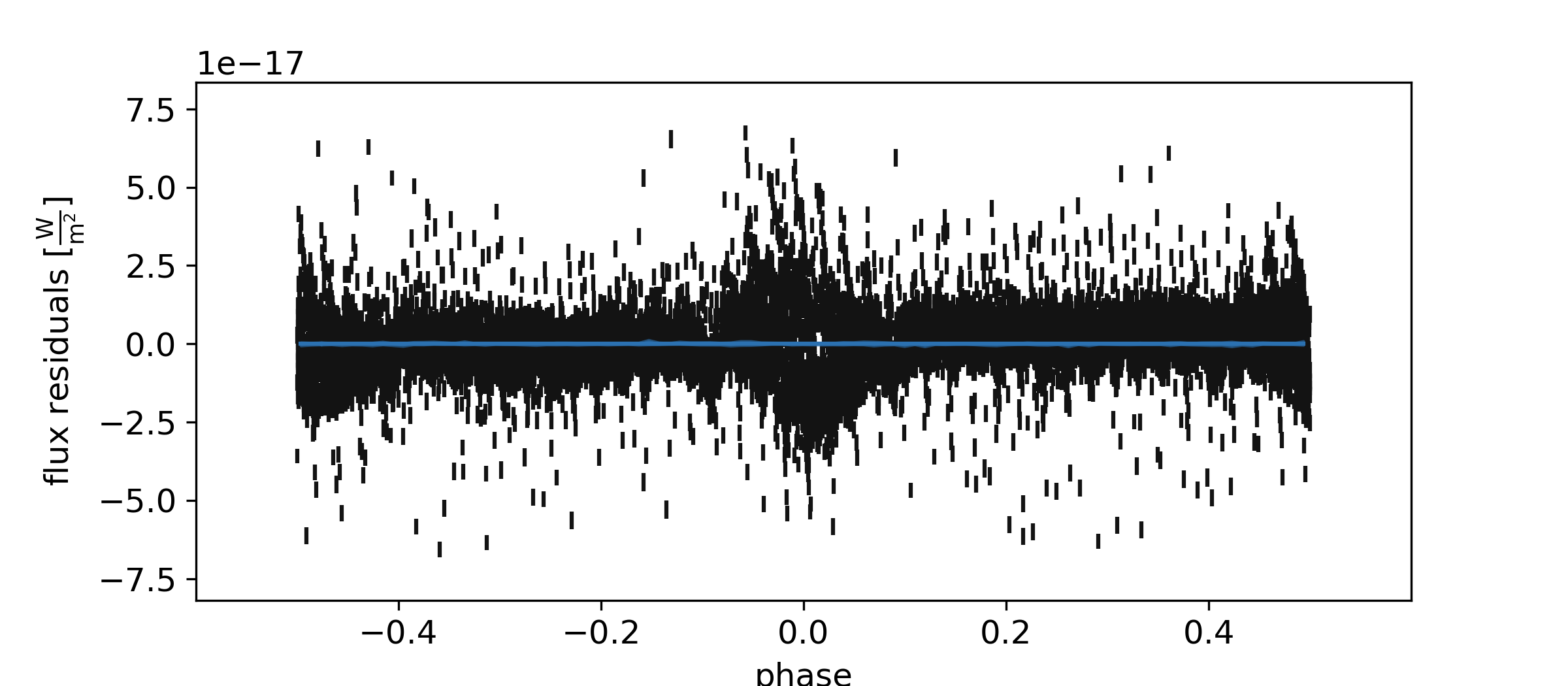}
    \caption{Phase-folded light curve of KIC 8314879 overplotted with 10 synthetic solutions drawn from the posterior distributions. Below, residuals between these synthetic models and the photometric data are plotted. There is some noticeable spread in the residuals near each eclipse. This may be resolved by increasing the number of points computed within the eclipse regions, or sampling over $e$ and $\omega$.}
    \label{fig:lc_8314879}
\end{figure}

A total of 15100 iterations were completed in the MCMC analysis of KIC 8314879. The average autocorrelation time for the parameters was estimated to be $1340\pm260$ iterations, where $\frac{T_2}{T_1}$ had the maximum autocorrelation time of 1540 iterations. 138900 more iterations would need to be performed to reach our benchmark for fair and unbiased sampling of 100 times the maximum autocorrelation time. The average acceptance fraction for this MCMC run was 0.0006.


Most MCMC walkers eventually converged to the same local minimum, however some of those spent thousands of iterations stuck in other local minima. A log-probability cutoff of 516555 was applied to center our results on the highest probability region, fully removing one walker from this solution and truncating the results of those walkers which took longer to converge. However, three walkers identified a different solution above the log-probability cutoff. This solution is not enclosed in the 1$\sigma$ widths of the posterior distributions, but can be seen on the fringes of Figure \ref{fig:cornerplots-8314879}. 

The light curve residuals presented in Figure \ref{fig:lc_8314879} exhibit minimal phase-space trends, however the spread of data points is larger near phases 0.0 and 0.5. The synthetic model has not quite captured the shape of the eclipse at these points, most likely due to the resolution of the forward model near these phases. If the spread is a result of limited resolution, it should be possible to reduce the spread by generating the forward model at more times during primary and secondary eclipse. If future testing shows that this change does not improve the residuals, then it may be necessary to sample over $e\sin\omega$ and $e\cos\omega$ instead of fixing these values at 0. The lack of phase-space trends indicates that any pulsations that exist in this star are unlikely to be tidally excited.  However, the scatter in the residuals remains relatively large, and may be due to pulsations of one or both components.  Future work will model pulsations of this system, using the binary constraints to narrow the possible parameter space.  The mass range we find here suggests this system likely contains at least one Slowly Pulsating B (SPB) star.

Parameter correlations uncovered by this MCMC analysis are more densely sampled than in the ellipsoidal variable, KIC 5957123. The one-dimensional posteriors of Figure \ref{fig:cornerplots-8314879} are uni-modal within 1$\sigma$, or even 2$\sigma$, of the median value, suggesting that most MCMC walkers have converged on a single solution. Computing more iterations will always provide a more precise estimate of the posterior distributions, but eventually the computing cost becomes unfeasible to manage. With limited computing resources, we must aim to strike a balance between time spent computing and relative improvement in parameter estimates. When deciding to conclude an MCMC run, we must rely on heuristic qualitative tests to evaluate convergence, as is no simple way to evaluate how long MCMC analysis must be performed to yield reliable results \citep{hogg_data_2018}. Since we have marginalized over appropriate nuisance parameters, and 60 of the 64 walkers converged to the same minimum after exploring the parameter space for 15100 iterations, we report parameter uncertainties from the widths of the posterior distributions. These uncertainties are displayed in Table \ref{tab:parameter-uncertainties}.


\begin{table}[tb]
\centering
	\caption{KIC 8314879 and KIC 10727668 Parameter Values} 
		\begin{tabular}{ccc}
	    	\hline
			\hline
			Parameter & KIC 8314879 & KIC 10727668 \\
			\hline
			\multicolumn{3}{c}{\textit{System Parameters}}\\ 
            $i\text{ }(^\circ)$ & $78.259^{+0.04}_{-0.016}$ & $79.33^{+0.3}_{-0.04}$\\
			$q$ & $0.689^{+0.004}_{-0.004}$ & $17.2^{+0.3}_{-0.3}$\\
			$a\text{ }(R_\odot)$ & $7.52^{+0.1}_{-0.14}$ & $11.1^{+0.3}_{-0.2}$\\
			$\frac{T_{\text{eff,2}}}{T_{\text{eff,1}}}$ & $0.753^{+0.005}_{-0.003}$ & $0.6501^{+0.0013}_{-0.002}$\\
			$\frac{R_2}{R_1}$ & $0.6710^{+0.0005}_{-0.0015}$ & $11.05^{+0.05}_{-0.04}$\\
			$\frac{R_1 + R_2}{a}$ & $0.5818^{+0.0002}_{-0.0002}$ & $0.2645^{+0.0004}_{-0.004}$\\
			$d\text{ }(\text{pc})$ & $8060^{+160}_{-200}$ & $5460^{+150}_{-130}$\\
			$e$ & ... & $0.018^{+0.002}_{-0.007}$\\
			$\omega\text{ }(^\circ)$ & ... & $88.9^{+0.98}_{-0.9}$\\
			\hline
			\multicolumn{3}{c}{\textit{Stellar Parameters -- Primary}}\\ 
			$T_{\text{eff,1}}\text{ }(\text{K})$ & $10660^{+78}_{-180}$ & $13580^{+50}_{-40}$\\
			$R_{\text{equiv,1}}\text{ }(R_\odot)$ & $2.62^{+0.04}_{-0.05}$ & $0.242^{+0.008}_{-0.006}$\\
            $M_1\text{ }(M_\odot)$ & $4.39^{+0.19}_{-0.2}$ & $0.189^{+0.011}_{-0.012}$\\
			Gr. Bright. & $0.933^{+0.02}_{-0.015}$ & $0.941^{+0.008}_{-0.009}$\\
			Refl. Frac. & $0.918^{+0.019}_{-0.012}$ & $0.89^{+0.05}_{-0.03}$\\
			\hline
			\multicolumn{3}{c}{\textit{Stellar Parameters -- Secondary}}\\ 
			$T_{\text{eff,2}}\text{ }(\text{K})$ & $8030^{+40}_{-80}$ & $8820^{+30}_{-30}$\\
			$R_{\text{equiv,2}}\text{ }(R_\odot)$ & $1.76^{+0.02}_{-0.03}$ & $2.68^{+0.07}_{-0.07}$\\
			$M_2\text{ }(M_\odot)$ & $3.03^{+0.12}_{-0.17}$ & $3.2^{+0.3}_{-0.2}$\\
			Gr. Bright. & $0.92^{+0.05}_{-0.09}$ & $0.943^{+0.009}_{-0.007}$\\
			Refl. Frac. & $0.830^{+0.011}_{-0.006}$ & $0.896^{+0.017}_{-0.009}$\\
			\hline
		\end{tabular}
		\label{tab:parameter-uncertainties}
\end{table}


\subsection{KIC 10727668}


\begin{figure*}
\centering
\begin{minipage}[b]{.45\textwidth}
    \centering
    \includegraphics[width=1.0\textwidth]{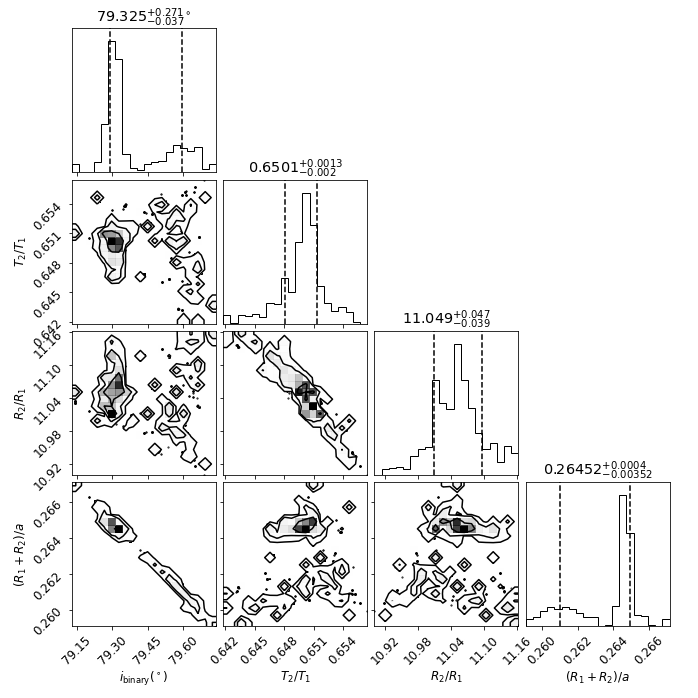}
\end{minipage}\qquad
\begin{minipage}[b]{.45\textwidth}
    \centering
    \includegraphics[width=1.0\textwidth]{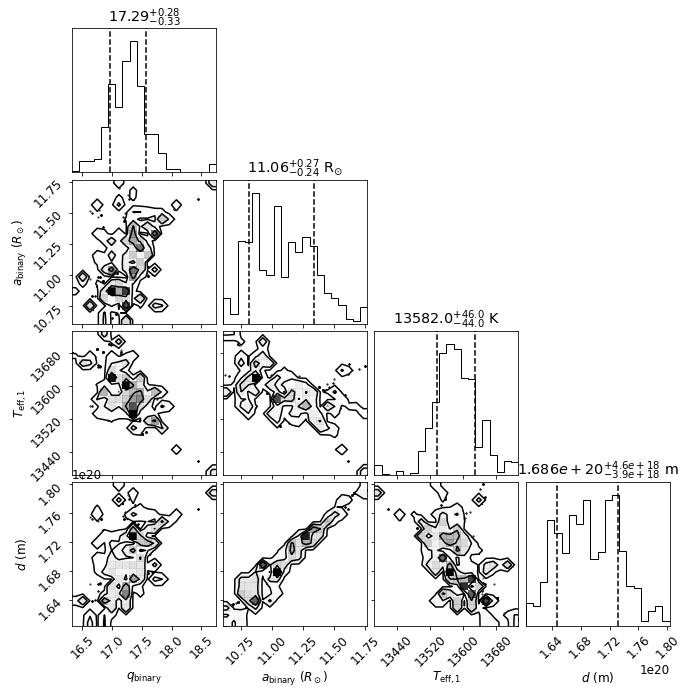}
\end{minipage}
\caption{Similar to Figure \ref{fig:cornerplots-5957123}, but for the posteriors uncovered by MCMC analysis of KIC 10727668. Three walkers with log-probability below 537075 were eliminated from this solution. Several posterior distributions are centralized and uni-modal, however $i$ and $\frac{R_1+R_2}{a}$ appear to have secondary modes which are included in the 1$\sigma$ spread.}\label{fig:cornerplots-10727668}
\end{figure*}

\begin{figure}[!t]
    \centering
    \includegraphics[width=0.45\textwidth]{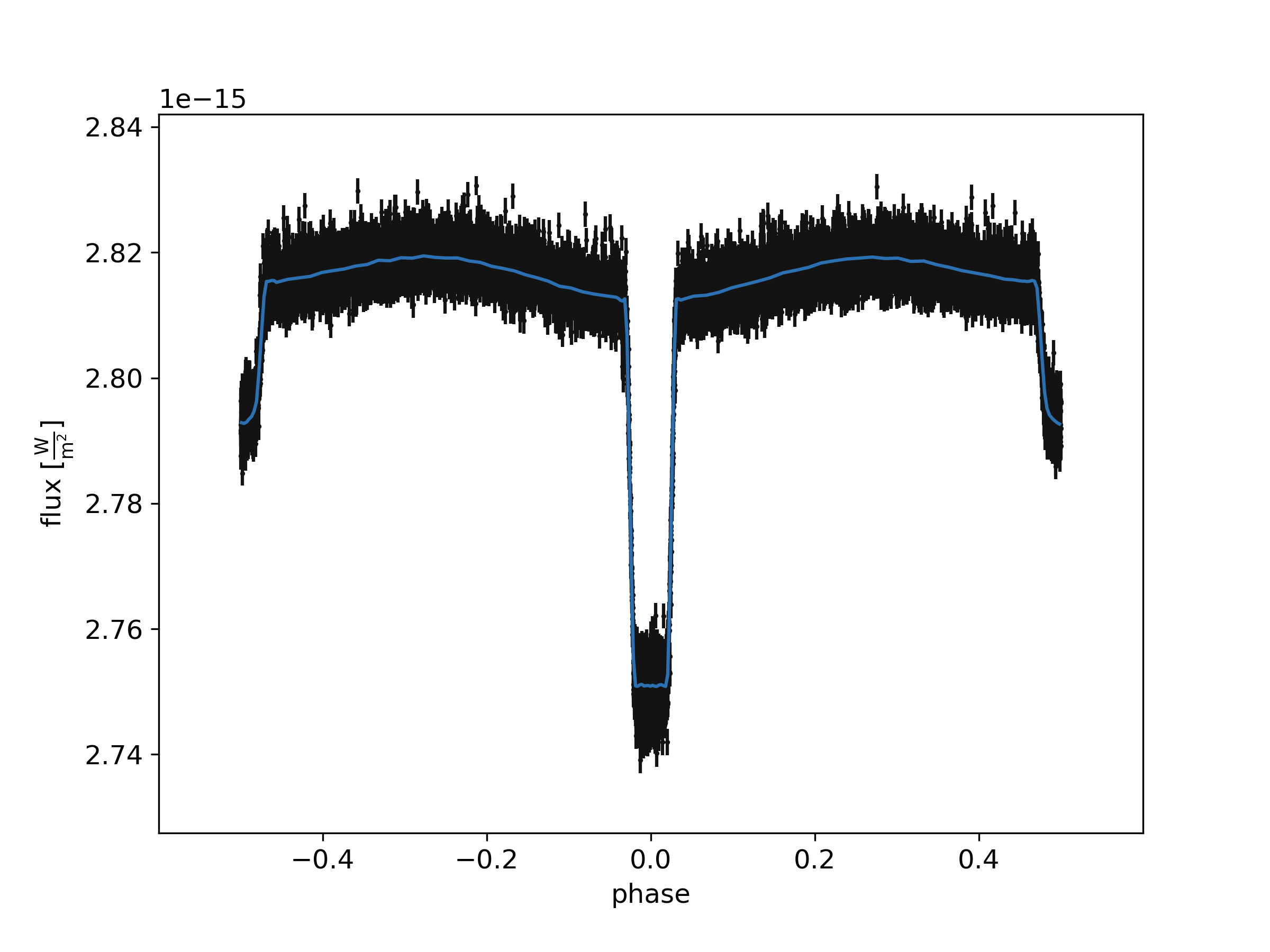}
    \includegraphics[width=0.45\textwidth]{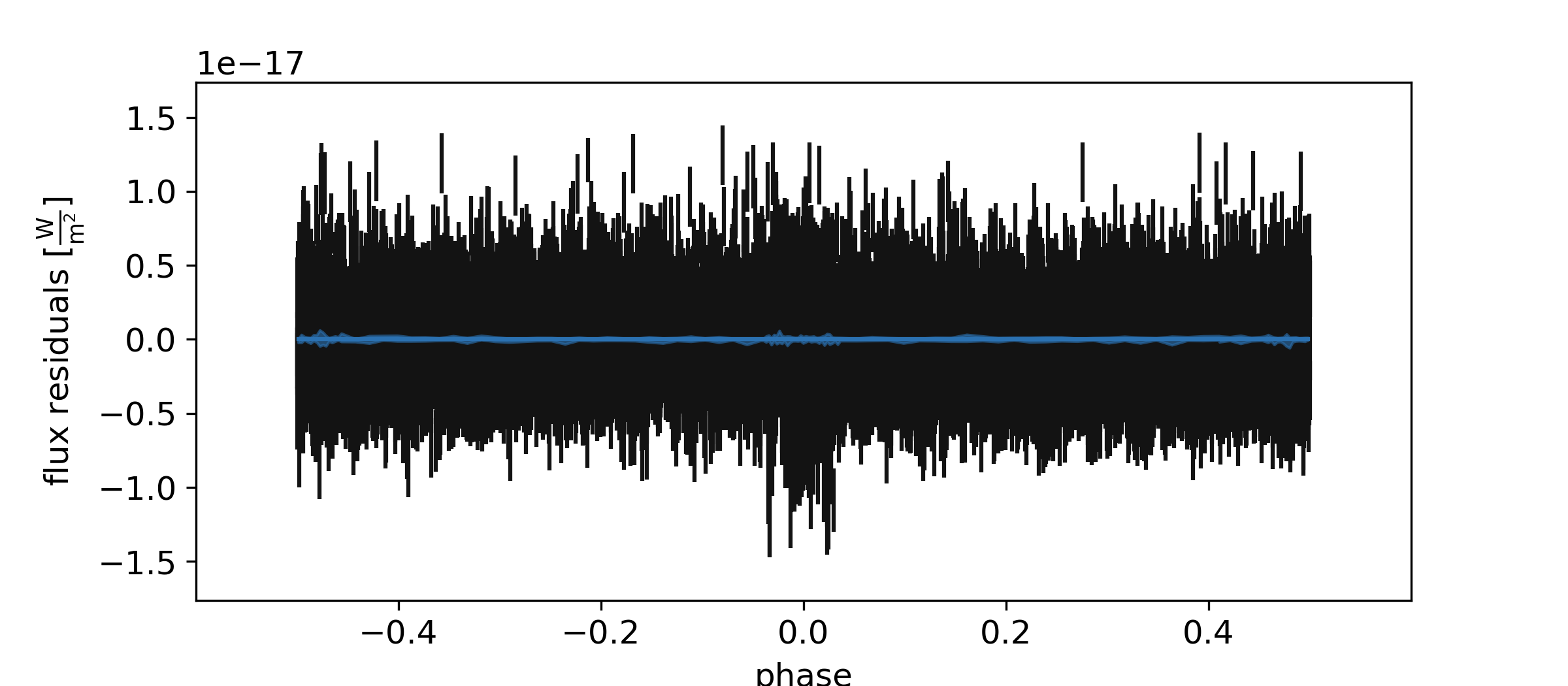}
    \caption{Phase-folded light curve of KIC 10727668 overplotted with 10 synthetic solutions drawn from the posterior distributions. Below, residuals between these synthetic models and the photometric data are plotted. These residuals exhibit little obvious trends in phase space, suggesting that the parameter sets drawn from the posteriors uncovered by our MCMC analysis suitably model this system.}
    \label{fig:lc_10727668}
\end{figure}

A total of 14200 iterations were completed in the MCMC analysis of KIC 10727668. The average autocorrelation time for the parameters was estimated to be $1150\pm300$ iterations, where $\frac{R_2}{R_1}$ had the maximum autocorrelation time of 1430 iterations. 128800 more iterations would need to be performed to achieve fair and unbiased sampling. The average acceptance fraction for this MCMC run was $4\times10^{-5}$.


For this solution, we adopted a log-probability cutoff of 537075, eliminating 3 walkers that were stuck in a lower probability region of the parameter space. This allows us to better resolve the two-dimensional Gaussian plots in the high-probability region around the proposed solution, shown in Figure \ref{fig:cornerplots-10727668}. The width of the $T_{\text{eff,1}}$ posterior distribution was consistently small throughout all runs, indicating that this system is much more sensitive to a change in temperature than the other binaries.

The light curve residuals displayed in Figure \ref{fig:lc_10727668} show no phase-space trends. This indicates that the synthetic solutions drawn from the posterior distributions of our MCMC analysis appropriately model the light curve variation of this system. Any deviation from this synthetic model appears to result from noise inherent in the data. Previous studies of this system suggest that one component may be a $\delta$-Scuti variable star \citep{wang_v1224_2018}, which is a likely contributor to the scatter seen throughout the residuals. Indeed, the Fourier spectrum for this system shows a small cluster of frequencies between 10 and 15 c d$^{-1}$, again consistent with a $\delta$ Scuti variable.  As for KIC 8314879, future work will involve modelling the pulsations of this system with the binary parameters found here as constraints.

Most posterior distributions are centralized and uni-modal, however some parameters seem to exhibit a lower-probability secondary mode, such as $i$ and $\frac{R_1+R_2}{a}$ in the left plot of Figure \ref{fig:cornerplots-10727668}. Walkers do not appear to pass from one mode to the other, suggesting that these secondary modes might be a consequence of stuck walkers which cannot find a path to the high probability solution, rather than a phenomenon of two distinct solutions. This seems a likely explanation given the rather low acceptance fraction for this system.

The posterior distributions for $e\sin\omega$ and $e\cos\omega$ were excluded from Figure \ref{fig:cornerplots-10727668}, since they did not form any significant correlations with other parameters. Rather, their posteriors appear to be independent and Gaussian, which can be seen in Figure \ref{fig:corner_binned_10727668}, the posterior distribution from the binned data set analysis. This behaviour, alongside the relatively uniform light curve residuals, suggests that we have likely sampled over appropriate nuisance parameters. Since our walkers have also converged to a solution which yields an appropriate fit to the photometric data, we may infer parameter uncertainties from the widths of the posterior distributions. These uncertainties are presented in Table \ref{tab:parameter-uncertainties}.


\section{Discussion} \label{sec:discussion}


This procedure was successful in obtaining robust parameter estimates for the two eclipsing binaries studied, but struggled to identify parameters and uncertainties for the ellipsoidal variable system. Throughout this section, we identify issues with this procedure and propose modifications for the benefit of future studies.

\subsection{Feasibility of this Analysis}\label{subsec:feasibility}

MCMC analysis was performed on binned light curve data to quickly identify the topology of the parameter space, and obtain initializing distributions for a full MCMC run. This significantly improved the effectiveness of the full data set run, since walkers converged quickly and were less likely to get stuck in local minima. Should this procedure be replicated, we strongly recommend that the model resolution be reduced for the initial binned data set run to encourage faster computation times. Larger per-point uncertainties and a smaller number of data points should permit a lower precision model to yield similar posterior distributions. This would permit quick initial sampling so that more compute resources may be allocated to process a higher resolution model in the full data set analysis. 

The stellar models used in this research struggle to account for noise in the \textit{Kepler} data, even when the noise-nuisance parameter $\sigma_{\text{lnf}}$ is implemented. Gaussian processes are supported by PHOEBE, but were not implemented in this analysis, as these options require that our light curve be computed over a range of times that spans the entire time series of the observed data. 
It is redundant to compute our stellar models for more than one period, since no intrinsic variability is modeled in the stars themselves. Applying Gaussian processes after interpolating from our one-period forward model to the entire time series could help account for stochastic noise without incurring longer computation times, and this could be investigated in future work on these stars.

One goal of this project was to examine the feasibility of establishing parameters for a larger set of \textit{Kepler} binaries using only time-series photometry. Considering that full MCMC analysis using PHOEBE models required roughly one CPU year of computation time, however, such a procedure is likely not feasible, especially considering the manual tuning required to optimize each MCMC run.


\subsubsection{Acceptance Fraction}\label{subsub:acceptancefraction}

\begin{figure*}
\centering
\begin{minipage}[b]{.45\textwidth}
    \centering
    \includegraphics[width=1.0\textwidth]{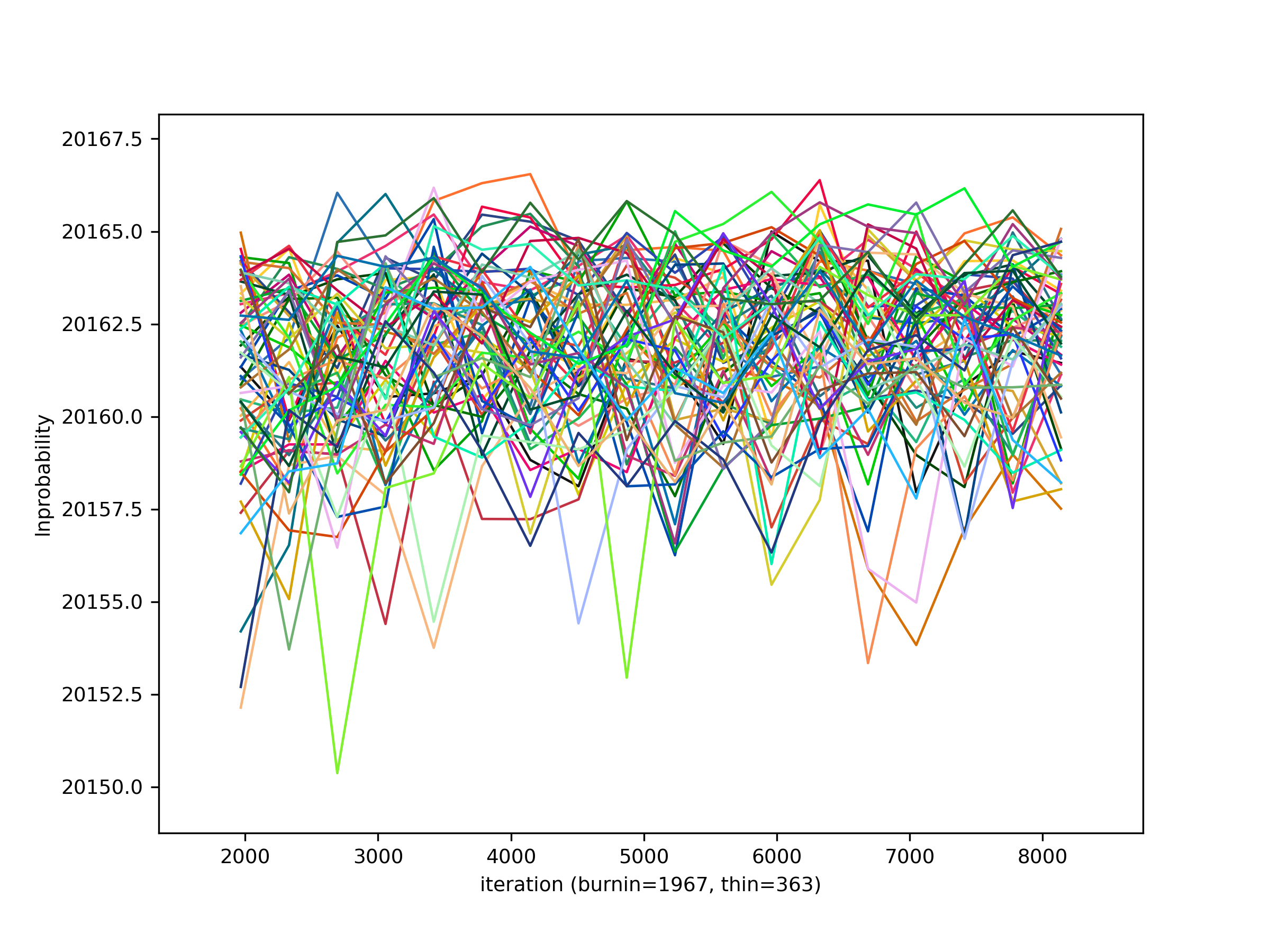}
\end{minipage}\qquad
\begin{minipage}[b]{.45\textwidth}
    \centering
    \includegraphics[width=1.0\textwidth]{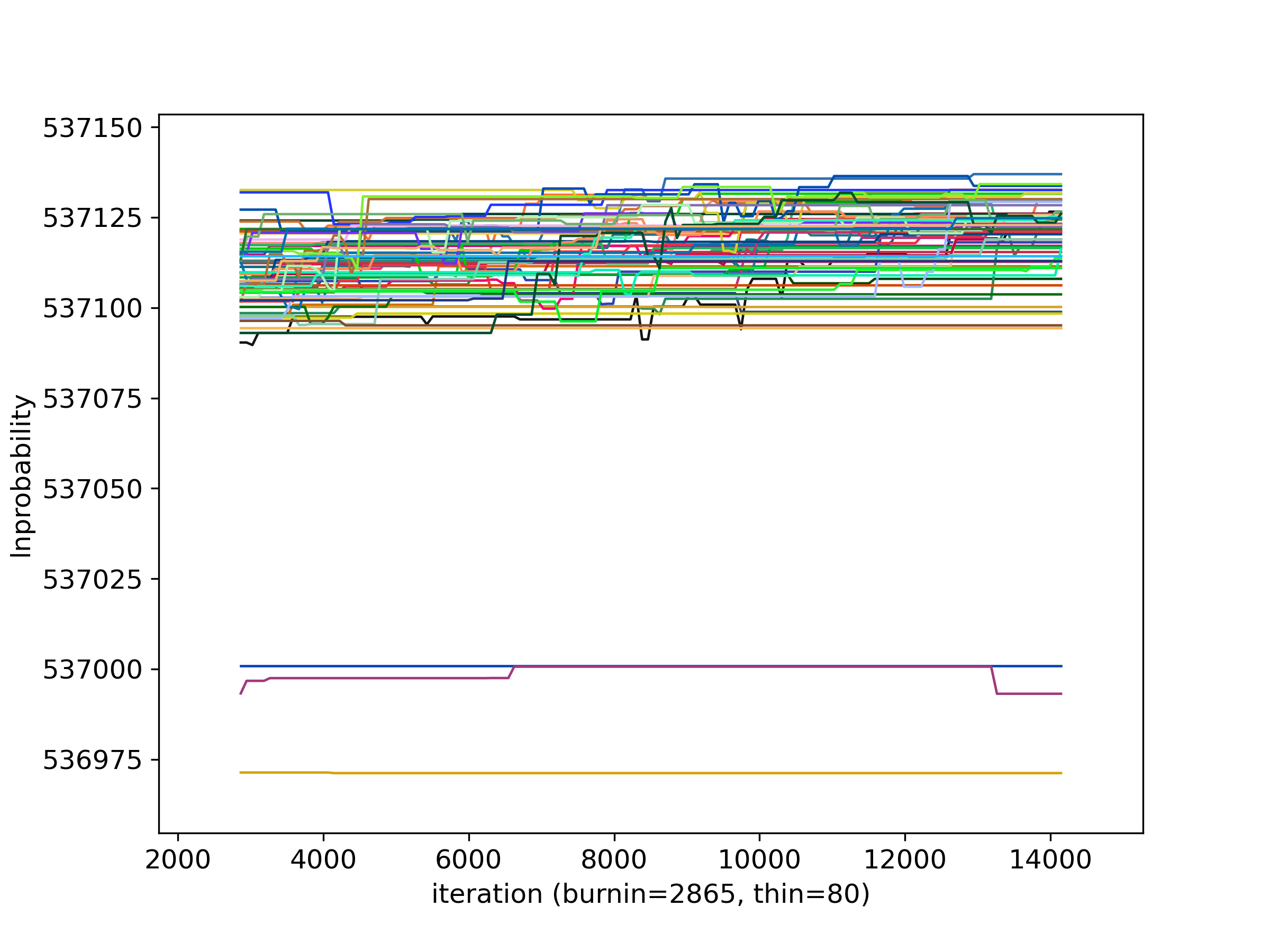}
\end{minipage}
\caption{Comparison of log-probability functions between the MCMC analyses of KIC 10727668, with burn-in and thinning applied. Each walker is represented with a curve of different colour. On the left is the log-probability function from the binned data set run. These chains are converged and readily accept new parameter combinations. On the right is the log-probability function from the full data set run. Notice that walkers in this run tend to become stuck -- they will spend many iterations in the same parameter combination, yielding unchanging log-probability.} 
\label{fig:log-probability_comparison}
\end{figure*}

The acceptance fraction of all three full data set MCMC runs was very low, with values of 0.0001, 0.0006, $4\times10^{-5}$ for KIC 5957123, KIC 8314879, and KIC 10727668 respectively. This is much smaller than the acceptance fraction from MCMC analysis of the binned data, which had acceptance fractions of 0.002, 0.003, and 0.008 respectively. A comparison of log-probability functions in Figure \ref{fig:log-probability_comparison} shows the differing behaviours of walkers in the binned data set analysis versus the full data set analysis, notably their tendency to accept new positions near the global minimum. Although the initial analysis with binned data successfully promotes sampling of the parameter space and provides a useful (albeit not robust) estimate of the posterior distribution, it does not directly encourage sampling in the subsequent full data set analysis. Rather, it only prevents initialization of walkers in a low-probability region of the parameter space.

\citet{hogg_data_2018} suggest that the average acceptance fraction can be improved by reducing the step size (the distance between the current position in parameter space and the proposed point) for the MCMC walkers, or by decreasing the variance of the proposal distribution. We found many of our models to be very sensitive to certain parameters ($d$, $T_{\text{eff,1}}$, $a$) which makes sampling for their correlations difficult. We suggest that the step size of emcee be reduced in any further analysis to encourage more appropriate sampling of these parameters. 


\subsection{Ellipsoidal Variables}

Ellipsoidal variables (ELVs) do not exhibit distinct eclipses in their light curves, and ellipsoidal variations cannot be decoupled from heating or reflection effects. As such, ELVs suffer from a lack of RV measurements far more than eclipsing binaries (EBs), since values for $i$ and $\frac{R_1+R_2}{a}$ cannot be measured directly from their light curves. Using PHOEBE's robust treatment of Roche effects and irradiation alongside MCMC sampling, we expected to produce parameter correlations for the ellipsoidal variable KIC 5957123, albeit to less precision than we could otherwise achieve for an EB.

It was difficult to propose an adequate initial solution for this system, and based on our results, it appears that our MCMC analysis could not find a suitable solution for KIC 5957123. Therefore, we must consider the possibility that our model does not adequately represent this system. Notably, our model does not fit the two peaks of differing height; other researchers have had success fitting similar features by implementing star spots \citep{negmeldin_physical_2019}. There is little prior information available to suggest that either star in this system has spots, nor where such spots should be placed. However, spots are relatively common in F-type main sequence stars, and Doppler imaging techniques have been used to identify the properties and evolution of star spots in other \textit{Kepler} binaries \citep{wang_kic_2021}. If spots were to be introduced to this PHOEBE model, then the spot parameters (temperature reduction, radius, latitude, and longitude) should be sampled in the MCMC analysis. 

If we had stellar spectra, and subsequently radial velocity measurements, we could measure $a\sin{i}$, $M_1$, $M_2$ (and by extension $q$), $V_\gamma$ (the systemic velocity), $T_{\text{eff,1}}$, and $T_{\text{eff,2}}$. Since this is not an eclipsing binary, we do not have constraints on the sum of fractional radii, however by using the distance estimate from Gaia we may constrain $R_1$ and $R_2$ and potentially construct a satisfactory stellar model. With such constraints, it might be possible to identify which phenomena is causing the difference in peak brightness at $\phi=0.25$ and $\phi=0.75$.

Without spectra, however, we struggle to confirm that our model adequately describes the data. Failing that implies our MCMC analysis of this system is not mathematically sound. We cannot claim that our walkers have adequately sampled the parameter space in the region around the true solution, and so despite our marginalization efforts we cannot quote the posterior widths as parameter uncertainties for this system.



\subsection{Eclipsing Binaries}

The method presented in this paper proved rather successful for establishing parameter correlations and values for eclipsing binaries. This is to be expected, as the geometry of an eclipsing binary's light curve provides information about several system parameters and ratios \citep{prsa_artificial_2008, southworth_binary_2020}. If steps are taken to appropriately marginalize (read: sample) over all nuisance parameters, it is possible to construct an appropriate model for an eclipsing binary system using only photometric data and distance measurements.

This method required roughly 10,000 iterations of MCMC sampling of the binned data, followed by roughly 14,000 iterations of MCMC sampling over the full data set. For the 64 MCMC walkers, this required a total of roughly 2 core years per system, not accounting for any of the failed runs or trial runs performed while refining this procedure! Therefore, this method is not possible without access to a high performance computing cluster. Successful sampling requires much computing time and human intervention. 

Other binary modelling software (ellc, PHOEBE Legacy) will run faster than PHOEBE 2.3.41, but with simplified physics. For this procedure, we wanted to use the most robust physical model possible, especially when trying to marginalize over nuisance parameters which may be constrained by second-order effects. For instance, we can constrain $q$ by the Roche-Lobe deformation that causes ellipsoidal variation in the out-of-eclipse regions of the light curves. This would not be possible in software which approximates stellar surfaces as spherical, and may be inaccurate in those which implement simplified potential surfaces. In order to appropriately marginalize over all nuisance parameters, the most thorough physical model was used at the cost of computation time. 

In summary, this method of MCMC analysis works for eclipsing binaries, but requires extensive computing resources and management. 
One of the main draws of studying eclipsing binary stars to constrain investigations of astroseismology, in which case, so long as the light curve data is treated ahead of time and the same treated data is used across both projects, the posterior distributions obtained from MCMC analysis may be used as prior distributions on astroseismological models.


\subsection{Stellar Pulsations}

At least two of the systems investigated here show some evidence of pulsations.  KIC 8314879 shows relatively large residuals, and the mass found here is consistent with a SPB variable.  B stars have convective cores, and the $g-$mode pulsations typical of SPB variables can be used to probe the convective overshoot \citep[e.g.][]{Papics2017}. With the addition of constraints from the binary orbit of the system, it will be possible to use the observed pulsations to find tight constraints on the behaviour of interior convection in this star.  

KIC 10727668 has been previously proposed to contain a $\delta$ Scuti variable \citep{wang_v1224_2018}, and that is borne out in our analysis.  Based on binary fitting, we find the mass of the secondary to be $3.2^{+0.3}_{-0.2}$ M$_{\odot}$, which is slightly higher than the typical range of masses seen in $\delta$ Scuti stars (1.4-2.5 M$_{\odot}$).  In the Fourier spectrum of this star, we see a cluster of frequencies between 10 and 15 c d$^{-1}$, which are likely to be $p-$mode pulsations.  As for KIC 8314879, these pulsation frequencies can be used to constrain the interior structure, including convective overshoot above the core.  This analysis will be the subject of future work.

\subsection{Concerning the Treatment of Data}

\subsubsection{Method of Flux Calibration}

Although our method of flux calibration appears to have yielded realistic stellar models, a more precise estimate of the \textit{Kepler} signal to flux conversion may exist elsewhere, and each CCD may require a separate calibration. In the raw data, each quarter had a different median flux -- the \textit{Kepler} spacecraft was rotated between quarters, so light from our binaries are detected by a different CCD each quarter, likely contributing to this flux discrepancy. A more robust method to de-trend our data might include calibrating each CCD separately, rather than applying one calibration and averaging the median flux of each quarter. An even better method may be to instead analyze each quarter separately. This would lead to fewer data points and less noise in each run, and allows for the flux of each quarter to be calibrated independently.


\subsubsection{Parameter Uncertainties}\label{subsec:parameter_uncertainties}




The widths of the posterior distributions returned by an MCMC run are highly dependent on the per-point uncertainties \citep{hogg_data_2010}, and to a lesser extent the number of data points used in our calculation of $\chi^2$. Binning the light curve data effectively involves transforming a data set of 13,500 points into a data set of 500 points. The uncertainty on each point is then defined to be the standard deviation of all points in each bin. For all three binaries, these per-point uncertainties were larger than the per-point uncertainties of the points from the Kepler light curve. It would be erroneous to claim that the posterior widths from MCMC analysis on these 500 points with large per-point uncertainties appropriately describes the parameter uncertainties for the binary system which produced those 13,500 points with smaller per-point uncertainties. Binning the light curve as done in this procedure encourages convergence and exploration of the parameter space. This is useful, especially as initial MCMC trial runs with a full data set did not converge in a reasonable time, and did a poor job exploring the parameter space, even with the inclusion of the noise-nuisance parameter. However, binning the light curve also artificially increases the widths of the posterior distributions, which would lead us to claim larger uncertainties than the data should allow.

The binning of light curve data has its place and its uses. For instance, it eliminates artificial instrumental noise that is not characteristic of the intrinsic variability of the binary, with less risk of eliminating intrinsic variability such as ellipsoidal variation in the out-of-eclipse regions. 
However, binning light curve data should not be used when attempting to obtain robust parameter uncertainties through MCMC analysis. Instead, the stellar model should be manipulated whenever possible to account for the behaviour of the data set, such as through the implementation of Gaussian processes or the noise-nuisance parameter (as discussed in section \ref{subsec:likelihood-function}). We acknowledge, however, that working with \textit{Kepler} data in this manner often proves difficult. Underestimated per-point uncertainty and high instrumental noise deter convergence and yield unreasonably small posterior widths, mainly because the log-probability function becomes so sensitive that walkers struggle to adopt new positions while exploring the parameter space.

Starting with MCMC analysis of a binned data set and using the resulting posteriors as an initializing distribution for a full data set run ensures that all walkers are initialized in a high probability region of the parameter space. However, due to the small acceptance fractions and large autocorrelation times characteristic of a full data set run, the full run still requires far more computational time and many more walkers than we would like in order to adequately sample the high-probability region around the global minimum.

\section{Conclusions} \label{sec:conclusions}

Using the PHOEBE stellar modeling package, we have constructed computational models of three \textit{Kepler} binary stars: KIC 5957123, KIC 8314879, and KIC 10727668. By applying MCMC methods to these models, we obtained estimates of the posterior probability distributions of the parameter space for each binary. From these posterior distributions, we attempted to justify our parameter estimates and extract uncertainties on each parameter. These posterior distributions also enabled us to extract light curve residuals for each binary.

This procedure proved relatively successful for the eclipsing binaries KIC 8314879 and KIC 10727668, but struggled to produce an adequate model for the ellipsoidal variable system KIC 5957123. There may be some phenomenon present in this system which was not fully realized by our model. It may be necessary to obtain radial velocities for this system in order to apply further constraints and generate a physically appropriate model. 

In all models, MCMC walkers struggled to converge due to stochastic noise in the \textit{Kepler} data and inadequate tuning of sampling parameters such as step size. Although the noise-nuisance parameter can be sampled to account for underestimated uncertainties, it may be necessary to include a more thorough characterization of scatter to appropriately model noisy \textit{Kepler} data.

Unfortunately, the faint magnitude of these stars (m$_K = 15.1-16.3$) makes spectroscopic follow up extremely challenging, requiring time on large telescopes with high resolution spectroscopy.  For the KIC 5957123 system, the models presented here are likely as tightly constrained as possible without further observations.

Appropriate residuals were obtained for KIC 8314879 and KIC 10727668, from which we may begin to probe pulsation characteristics in further analysis.
Both of these systems show signs of residual pulsation, possibly associated with SPB and $\delta$ Scuti variables respectively.   Asteroseismic modelling has been shown to be effective at constraining the internal structure of stars, including the internal rotation profile and convective core overshoot \citep[e.g.,][]{Papics2017, pedersen_shape_2018}.  Future work on these systems will focus on improving the binary models presented here to provide tight constraints on the stellar properties. The addition of the constraints provided from the binary modelling will enable us to place tighter constraints on the internal structure than would be possible with asteroseismology alone.




\begin{acknowledgments}
This research was funded by resources from NSERC. The authors would also like to thank Compute Canada, the PHOEBE team, the MAST archive, and the Kepler GO program. The authors are grateful to the anonymous referee for their helpful comments. This research has made use of NASA's Astrophysics Data System.
\end{acknowledgments}

%

\vspace{5mm}
\facilities{\textit{Kepler}, Gaia, SORCE
}


\software{astropy \citep{2013A&A...558A..33A,2018AJ....156..123A},  
          PHOEBE \citep{prsa_computational_2005}, 
          emcee \citep{foreman-mackey_emcee_2013}, numpy \citep{harris2020array}
          }

\appendix

\section{Posterior Distribution Corner Plots}
\label{appendix:cornerplots}

Here we display the full corner plots from our MCMC runs, representing all parameters sampled. We have excluded the gravity brightening and reflection fraction parameters from these corner plots, as they produced uniform distributions which did not exhibit any correlation with the other parameters sampled.

\subsection{Corner Plots from MCMC Analysis of Binned data}
\label{appendix:binned_cornerplots}

\begin{figure}[h]
    \centering
    \includegraphics[width=0.9\textwidth]{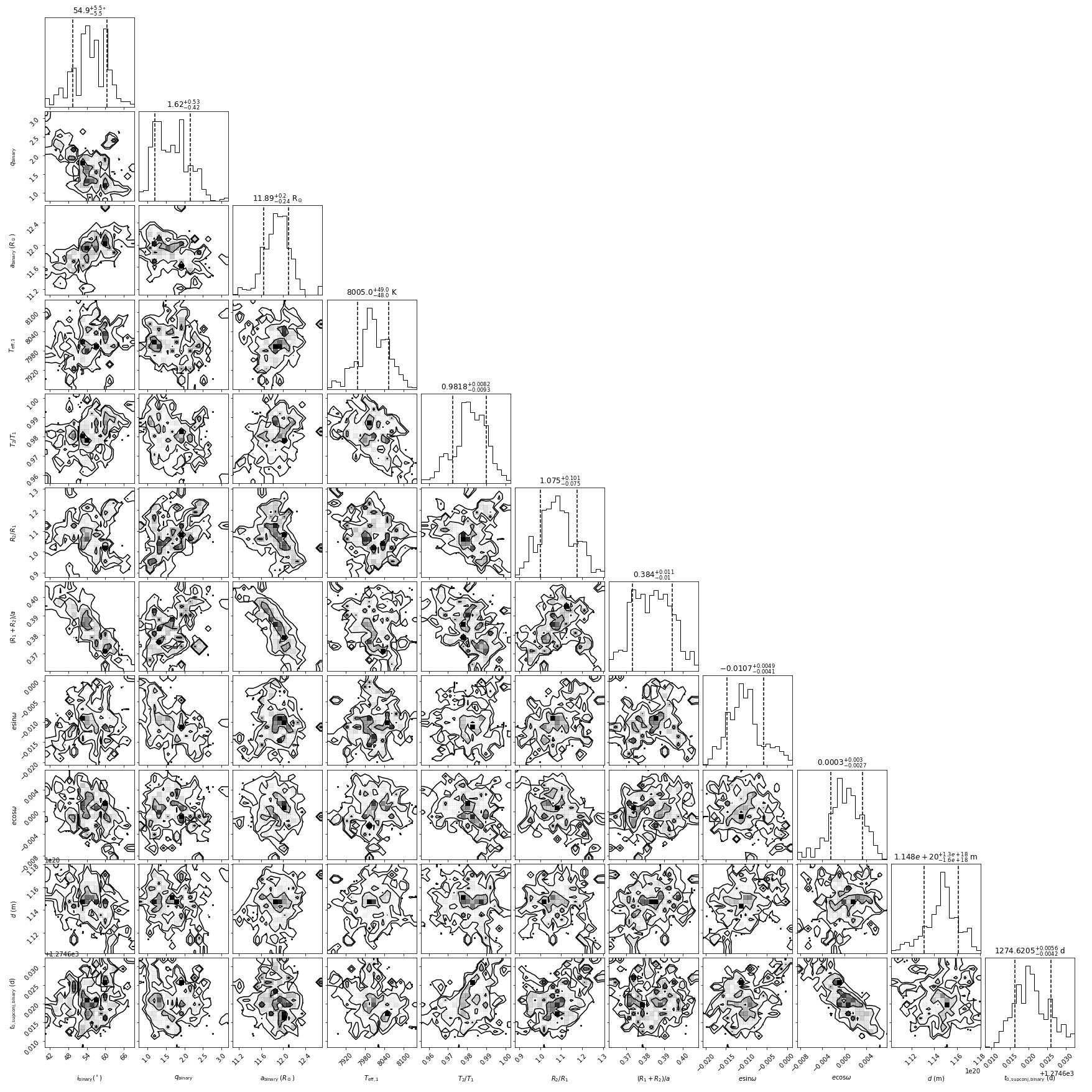}
    \caption{Full corner plot showing the estimated posterior distribution from MCMC analysis of the binned light curve data from KIC 5957123.}
    \label{fig:corner_binned_5957123}
\end{figure}

\begin{figure}[h]
    \centering
    \includegraphics[width=1.0\textwidth]{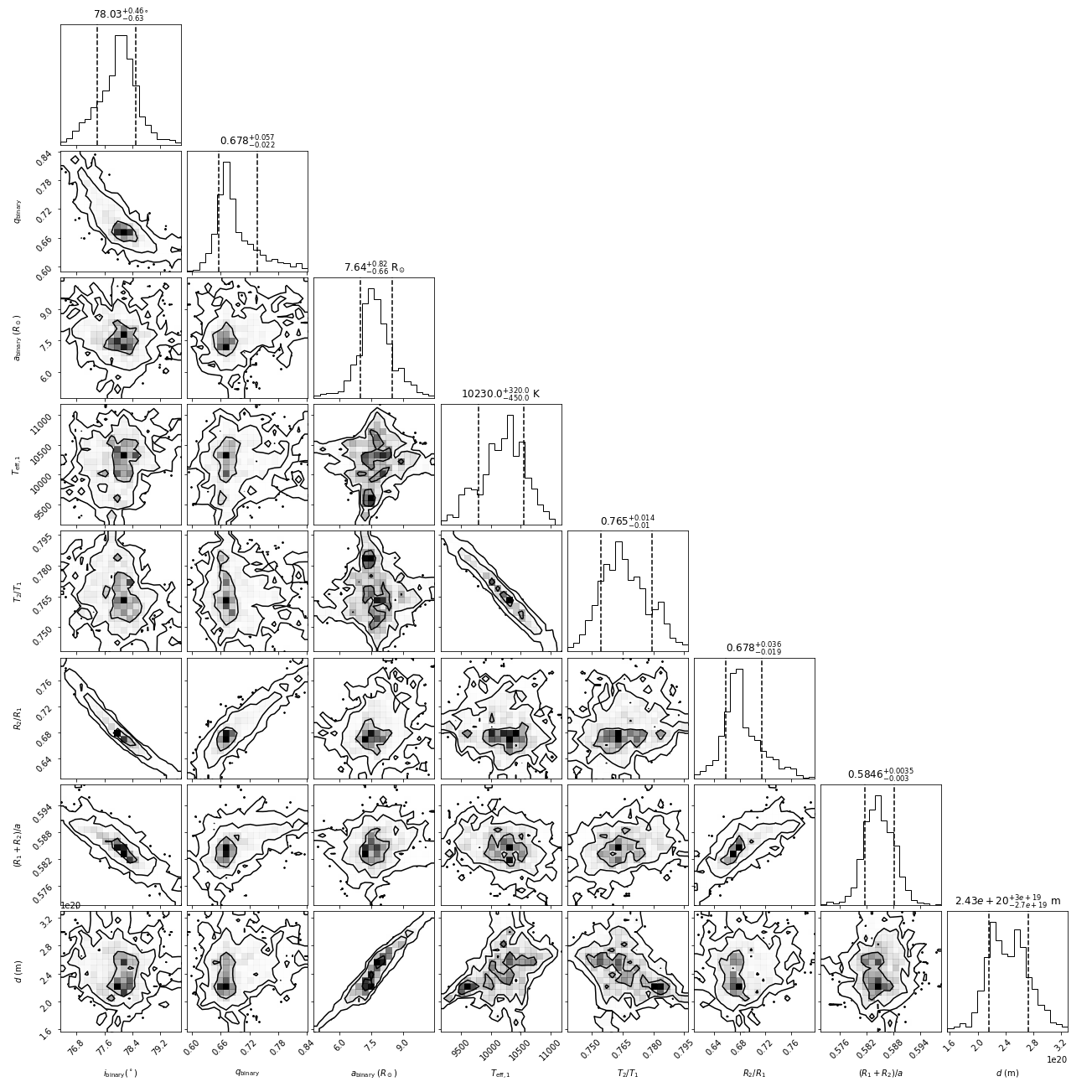}
    \caption{Full corner plot showing the estimated posterior distribution from MCMC analysis of the binned light curve data from KIC 8314879.}
    \label{fig:corner_binned_8314879}
\end{figure}

\begin{figure}[h]
    \centering
    \includegraphics[width=1.0\textwidth]{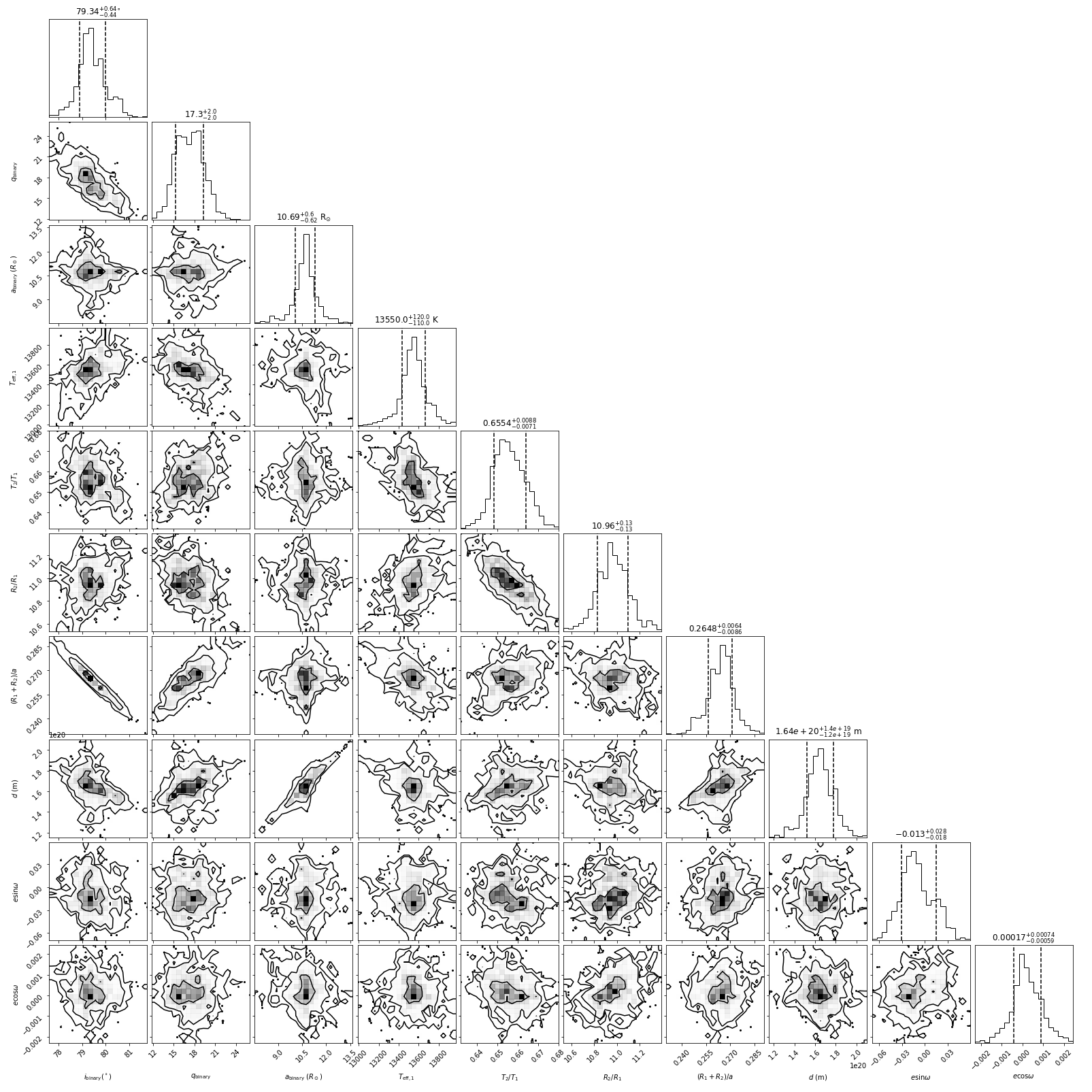}
    \caption{Full corner plot showing the estimated posterior distribution from MCMC analysis of the binned light curve data from KIC 10727668.}
    \label{fig:corner_binned_10727668}
\end{figure}

Figures \ref{fig:corner_binned_5957123}, \ref{fig:corner_binned_8314879}, and \ref{fig:corner_binned_10727668} show the posteriors for the binned data set runs of KIC 5957123, KIC 8314879, and KIC 10727668 respectively.

\newpage
\subsection{Corner Plots from MCMC Analysis of Full data}
\label{appendix:full_cornerplots}

\begin{figure}[h]
    \centering
    \includegraphics[width=1.0\textwidth]{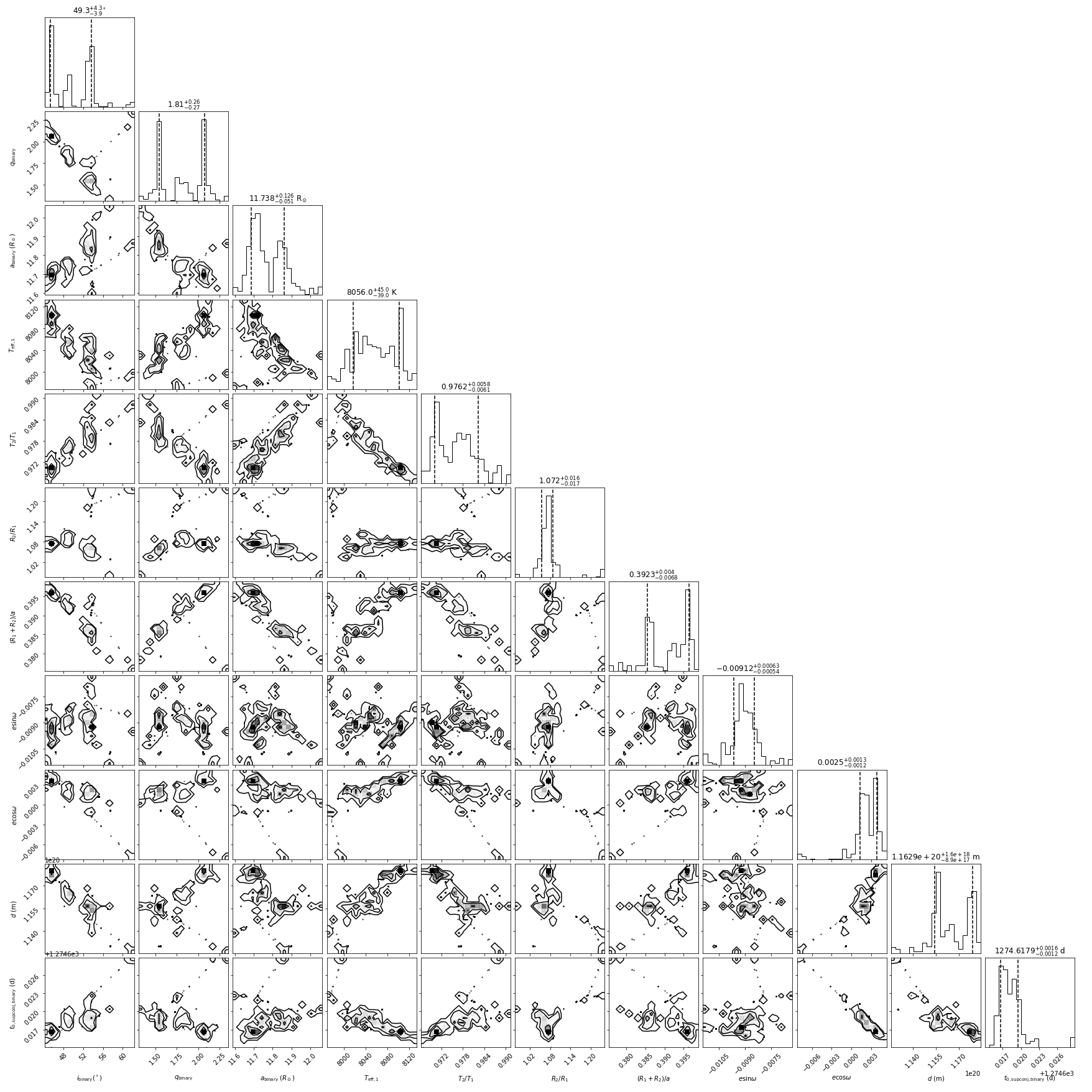}
    \caption{Full corner plot showing the estimated posterior distribution from MCMC analysis of KIC 5957123.}
    \label{fig:corner_full_5957123}
\end{figure}

\begin{figure}[h]
    \centering
    \includegraphics[width=1.0\textwidth]{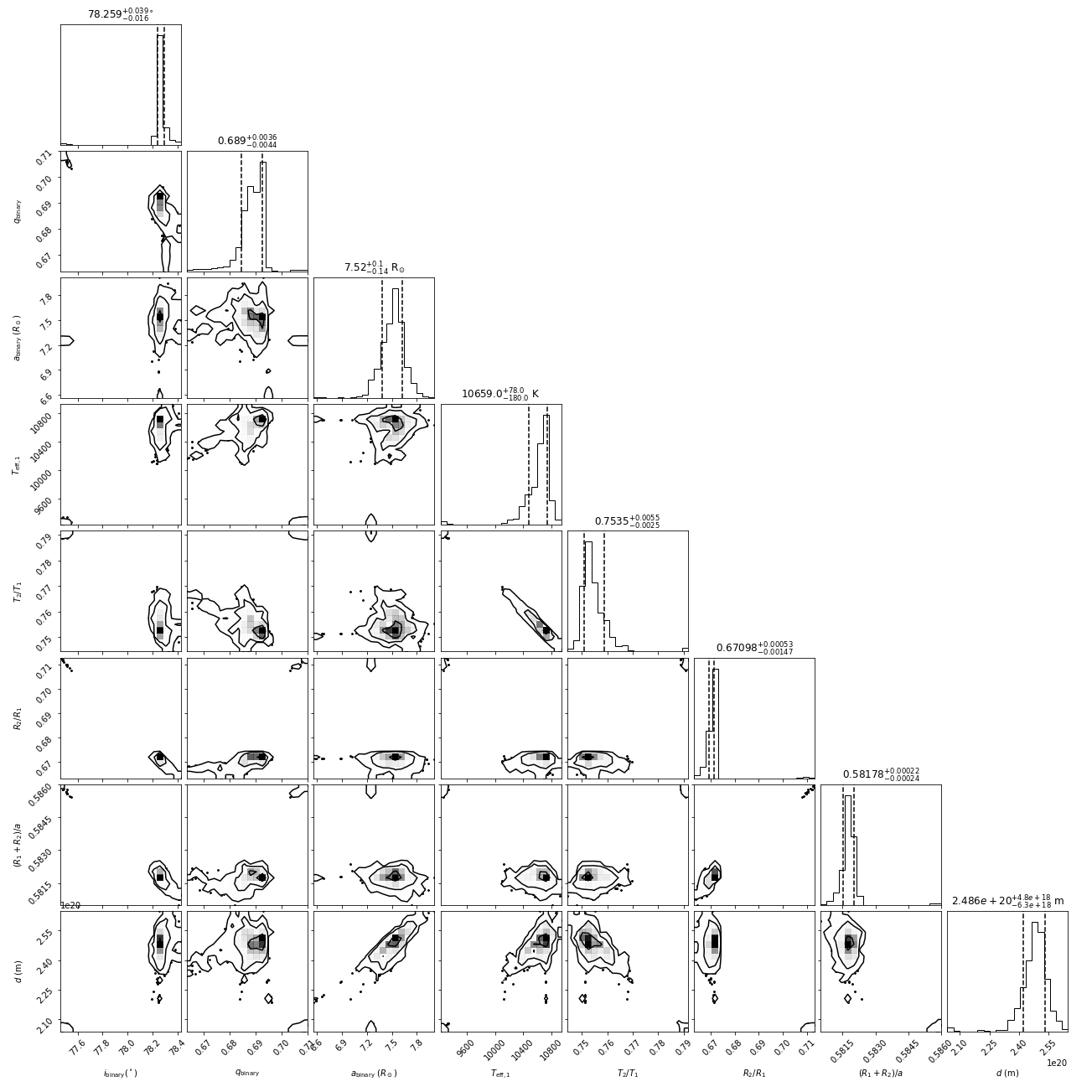}
    \caption{Full corner plot showing the estimated posterior distribution from MCMC analysis of KIC 8314879.}
    \label{fig:corner_full_8314879}
\end{figure}

\begin{figure}[h]
    \centering
    \includegraphics[width=1.0\textwidth]{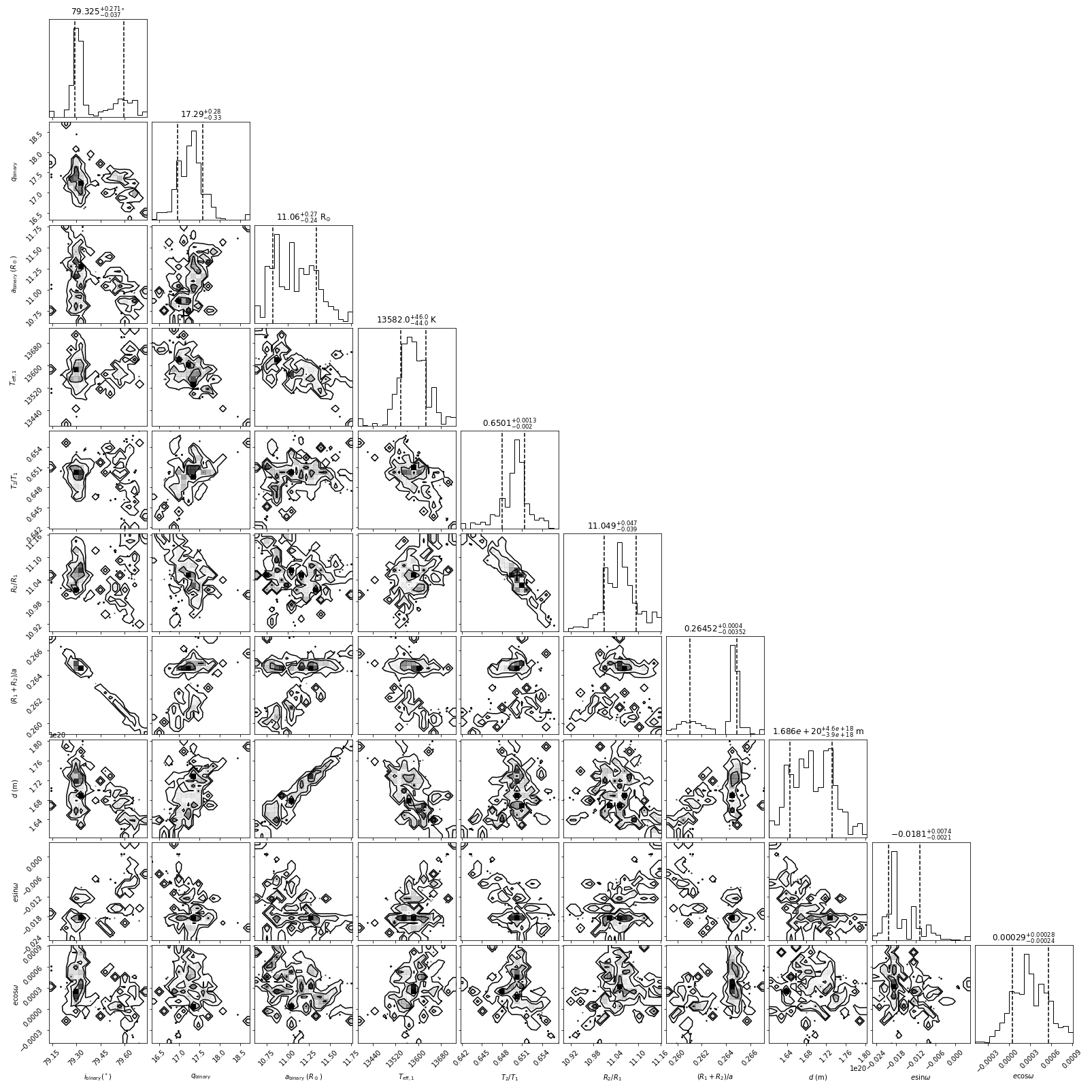}
    \caption{Full corner plot showing the estimated posterior distribution from MCMC analysis of KIC 10727668.}
    \label{fig:corner_full_10727668}
\end{figure}

Figures \ref{fig:corner_full_5957123}, \ref{fig:corner_full_8314879}, and \ref{fig:corner_full_10727668} show the posteriors for the full data set analysis of KIC 5957123, KIC 8314879, and KIC 10727668 respectively. Again, gravity darkening and reflection fractions are excluded from these plots, as is the noise-nuisance parameter.

\newpage
\bibliography{scientificpaper}{}
\bibliographystyle{aasjournal}



\end{document}